\documentclass[preprintnumber,trackchanges]{aastex701}
\usepackage{ae,aecompl}
\usepackage{graphicx}
\usepackage{amsmath}
\usepackage{amssymb}
\usepackage{supertabular}
\usepackage{hyperref}
\usepackage{xcolor}
\usepackage{tabularx}
\usepackage{booktabs}
\usepackage{bm}
\usepackage{academicons}

\usepackage{hyperref}
\hypersetup{
    colorlinks=true,
    urlcolor=blue,
    linkcolor=blue,
    citecolor=blue
}

\def\bea{\begin{eqnarray}} \def\eea{\end{eqnarray}}
\def\fm3{\;\text{fm}^{-3}}

\newcommand{\D}{{\rm d}}

\newcommand{\ie}{i.e.,~}
\newcommand{\eg}{e.g.,~}

\begin{document}

\title{Reconstruction of fast-rotating neutron star observables with the neural network}

\author{Wen Liu}
\affiliation{College of Artificial Intelligence, Suzhou Chien-Shiung Institute of Technology, Suzhou, Jiangsu 215411, China}
\email[]{wenliu0909@outlook.com}

\author[orcid=0000-0003-3757-3403]{Lingxiao Wang}
\affiliation{RIKEN Interdisciplinary Theoretical and Mathematical Sciences (iTHEMS), Wako, Saitama 351-0198, Japan}
\affiliation{Institute for Physics of Intelligence, The University of Tokyo, Hongo,  Tokyo 113-0033, Japan}
\email[show]{lingxiao.wang@riken.jp}

\author[orcid=0000-0001-9189-860X]{Zhenyu Zhu}
\affiliation{Center for Computational Relativity and Gravitation, Rochester Institute of Technology, Rochester, NY 14623, USA;}
\affiliation{Tsung-Dao Lee Institute, Shanghai Jiao Tong University, Shanghai, 201210, China}
\email[show]{zhenyu.zhu@rit.edu}

\date{\today}

\begin{abstract}
Rotation can significantly affect neutron-star (NS) properties, but accurate modeling of rapidly rotating NSs requires solving a two-dimensional, axially symmetric system, making traditional calculations too expensive for inference analyses that demand a large amount of model evaluations. We develop a causal convolutional neural networks that preserve the chronological-like dependence of NS properties on the equation of state (EoS) and rapidly reconstruct observables for static, Keplerian, and rotating configurations. Using \texttt{RNS}, we generate a dataset of NS observables and use it to train our networks. We validate our networks with three representative EoS (SFHo, SLy4, and DD2) and find that the they accurately reproduce the \texttt{RNS} results. The trained networks evaluate NS configurations for a single EoS in $\sim 50$ms, providing a substantial speedup over typical \texttt{RNS} runtimes of $\sim 30$ min and enabling efficient inference analyses involving rapidly rotating NSs.
\end{abstract}

\section{Introduction}
\label{sec:intro}

\setcounter{footnote}{0}

The recent progress in gravitational wave (GW) detection by the LVK Scientific Collaboration~\citep{2017PhRvL.119p1101A, 2019PhRvX...9a1001A, 2018PhRvL.121p1101A, 2020ApJ...892L...3A} and in the mass-radius measurements with the NICER mission~\citep{2019ApJ...887L..24M, 2019ApJ...887L..21R, 2021ApJ...918L..28M, 2021ApJ...918L..27R, 2024ApJ...971L..20C} have significantly advanced our understanding of neutron stars (NSs) and their equation of state (EoS)~\citep{2017ApJ...850L..19M, 2018ApJ...852L..29R, 2018PhRvL.121p1101A, 2018ApJ...857L..23R, 2018PhRvL.120q2703A, 2018PhRvD..98f3020Z, 2018PhRvL.120z1103M, 2019ApJ...881...73W, 2019MNRAS.489L..91C, 2021MNRAS.505.1661B, 2022ApJ...926..196H, 2023PhRvC.108b5809Z, 2023ApJ...943..163Z}. The key NS observables, \ie the tidal deformability (TD) and the mass-radius relation, can be one-to-one mapped from the EoS through the TOV and tidal equations~\citep{2008PhRvD..77b1502F, 2008ApJ...677.1216H, 2010PhRvD..81l3016H, 2020PhRvD.102h4058Z}. In addition, the upcoming third-generation GW detectors, the Einstein Telescope (ET) and Cosmic Explorer~\citep{2010CQGra..27h4007P, 2012CQGra..29l4013S, 2019BAAS...51g..35R, 2020JCAP...03..050M}, are expected to detect more and higher-quality GW signals from NS sources, which will enable more accurate constraints of the EoS and NS properties~\citep{2020ApJS..250....6W, 2022arXiv220501182G, 2022PhRvD.106l3529G, 2022ApJ...941..208I, 2023PhRvD.108l2006I, 2024PhRvD.110d3013W, 2024PhRvD.109j3035H, 2025arXiv250811875Z}. Meanwhile, high-precision measurements of the NS moment of inertia (MOI) via pulsar timing are expected to become achievable in the coming years~\citep{2020MNRAS.497.3118H, 2021ApJ...915L..12F, 2021PhRvX..11d1050K}. Together, these rapid observational advances motivate robust and efficient methods to infer the NS EoS and related observables from increasingly precise multi-messenger data.

GW signals from binary neutron star (BNS) or black hole-neutron star (BHNS) systems encode the information about the masses and tidal deformabilities of neutron stars. However, the imprint of the TD on the waveform is weak~\citep{2019PhRvD.100d4003D} and can only be identified when the signal-to-noise-ratio (SNR) is sufficiently high. In practice, the primary method for distinguishing NSs from BHs in GW observation relies on their masses. This approach could fail when the component masses fall into the mass gap, where both neutron stars and black holes may exist in the same mass range. Consequently, fast-rotating neutron stars with very large masses have been proposed as the component of several GW sources~\citep{2020ApJ...902...38Z, 2021ApJ...908L..28N, 2025arXiv250808750M}. In addition, several fast-rotating pulsars have been observed~\citep{2006Sci...311.1901H, 2017ApJ...846L..20B, 2022ApJ...934L..17R}, and the effects of rotation cannot be neglected when inferring the EoS of NS~\citep{2025arXiv250209200W}.

The rotation of NSs can be treated as a perturbation to the background TOV solution when the NS spin is small~\citep{2001CQGra..18..969A}. However, for fast-rotating NSs with rotational rates close to their Keplerian limit, the effects of rotation on spacetime become comparable to those of the matter itself, and the perturbative assumption is no longer valid. In this scenario, a two-dimensional axially symmetric system has to be solved for a comprehensive and accurate description. Several codes have been developed to model such fast-rotating NSs~\citep{1989MNRAS.237..355K, 1994ApJ...422..227C, 1995ApJ...444..306S, 1998PhRvD..58j4020B, 1998A&AS..132..431N, 1999A&A...349..851G, 2021PhRvD.104b4057P, 2026arXiv260105176T}, and a rotating stellar configuration can be obtained once the EoS, central density, and angular velocity are specified. 

However, due to the significant computational expense of the two-dimensional calculations for rapidly rotating systems, such methods are impractical for analyses that require a large number of model evaluations with short computational times (\eg Bayesian inference). Consequently, the fitting formulas that effectively capture the impact of rotation are commonly adopted in the analyses~\citep{2016PhRvD..94h3010L, 2017ApJ...844...41L, 2025arXiv250209200W}. These fitting formulas usually only involve a few parameters and do not capture the EoS dependence with sufficiently high accuracy. Alternatively, a neural network can fulfill the requirements of both accuracy and computational speed if the training data are sufficient. By training the neural network with a large number of EoS and mass-radius-TD datasets, it has been shown that the neural network can learn and capture the essence of TOV equations, and can provide highly accurate predictions of mass, radius, and TD for a given EoS input~\citep{2018PhRvD..98b3019F, 2020PhRvD.101e4016F, 2020A&A...642A..78M, 2022JCAP...08..071S, 2023PhRvD.107h3028S, 2024ApJ...974..285R}.

To train a neural network that reconstructs the properties of fast-rotating NSs from the EoS, we first require a large number of physical and reasonable NS EoSs that can cover most of their possible variety. Indeed, many approaches for modeling or parameterizing the NS EoS have been developed to facilitate Bayesian inference of EoS using GW, mass-radius, and nuclear data. These includes various nuclear physics models~\citep{2018PhRvC..97c5805Z, 2018ApJ...862...98Z, 2023PhRvC.108b5809Z, 2023ApJ...943..163Z, 2020ApJ...897..165T,2023ApJ...943..163Z, 2025PhRvD.112f3003L, 2018PhRvC..97c5805Z, 2018ApJ...862...98Z, 2019PhRvC..99b5804Z, 2023PhRvC.108b5809Z, 2025PhRvD.112d3018T, 2018PhRvC..98e4618W, 2024PhRvC.109e4623W, 2025PhRvC.111e4605W}, Taylor expansions of nuclear matter and symmetry energy~\citep{2018ApJ...859...90Z, 2021ApJ...921..111Z, 2024PhRvD.110j3040L, 2025arXiv250415893W}, piecewise polytropes~\citep{2018PhRvL.120q2703A, 2018PhRvL.120z1103M, 2019ApJ...881...73W, 2025arXiv251205315S}, spectral decomposition~\citep{2010PhRvD..82j3011L, 2022PhRvD.105f3031L, 2022ApJ...926..196H, 2024PhRvD.110h3030L, 2024arXiv241014674Y, 2024arXiv240715753V}, Gaussian process~\citep{2018PhRvL.121p1101A, 2024PhRvD.109b3020L, 2025PhRvD.112f3003L, 2025CQGra..42t5008N, 2025PhRvD.112j3023F}, and speed-of-sound modelings~\citep{2021ApJ...919...11H, 2022ApJ...939L..34A, 2023ApJ...950...77H, 2023SciBu..68..913H, 2025PhRvD.111g4026L}. There is no perfect modeling for the EoS: some of them can provide a wide variety of sound-speed behaviors and EoS, while lacking the physical information of nuclear matter and NSs; others may incorporate the principles of nuclear physics and interactions, but the output EoS may be constrainted to some specific form. In the present scenario, however, a barotropic EoS parameterization that can cover most of the variety is preferable. Therefore, we adopt a speed-of-sound modelings with a feedforward neural network (FNN) method~\citep{2021ApJ...919...11H, 2023ApJ...950...77H, 2023SciBu..68..913H}, which in principle can reproduce all functional forms of sound speed.

This paper is organized as follows. In Sec.~\ref{sec:method}, we describe our methodology, including the FNN-based EoS generation, the \texttt{RNS} code for computing rotating NS observables, and the architecture of the causal convolutional neural networks. In Sec.~\ref{sec:training}, we describe the data cleaning procedure and present the training 
results of the three networks. In Sec.~\ref{sec:results}, we validate our networks against three representative EoSs (SFHo, SLy4, and DD2), and discuss the mass--radius and 
mass--angular-velocity relations as well as the interpolation procedure for obtaining configurations at fixed angular velocity. We summarize our conclusions in 
Sec.~\ref{sec:conclusion}. Additional results for other NS observables are presented in Appendix~\ref{sec:appendix1}.

\section{Methodology}
\label{sec:method}

\subsection{Equation of state}

We adopt the feedforward neural network (FNN) method to generate the core and inner-crust EoS of NS~\citep{2021ApJ...919...11H, 2023ApJ...950...77H, 2023SciBu..68..913H}. We recall that the squared sound speed of EoS can be written as
\begin{eqnarray}
  \label{eq:ffnn_eos}
  c_s^2 & = & \sum_{i=1}^{10} \omega_{2i} \sigma(\omega_{1i} \log \rho + b_i) + B,
\end{eqnarray}
where $\omega_{1i}$, $\omega_{2i}$, and $b_i$ are free parameters with values ranging from $-10$ to $10$, and $B$ is fixed by matching the sound speed of outer-crust BPS EoS~\citep{1971ApJ...170..299B}. The $\rho$ and $c_s^2$ represent the rest-mass density and the squared sound speed of NS matter, respectively. Most EoS randomly generated from Eq.~(\ref{eq:ffnn_eos}) are not physically acceptable if no constraints are imposed. We therefore require pressure $p$ at saturation density to satisfy $3.12 \times 10^{33} \leq p(\rho_{\rm sat}) \leq 4.7 \times 10^{33} {\rm dyn/cm^2}$, and we further set a lower bound on pressure at $1.85 \rho_{\rm sat}$, $p(1.85 \rho_{\rm sat}) \geq 1.21 \times 10^{34} {\rm dyn/cm^2}$.

Furthermore, we consider only nucleonic matter in NSs, for which a monotonic sound speed is expected. Accordingly, we also impose a monotonicity constraint when generating the EoSs. In addition, we rule out soft EoS that cannot produce a maximum NS mass larger than $1.4\,M_\odot$. After applying these constraints, the generated EoSs benefit from both large variety and reliability. We generate $20{,}000$ EoSs and input them into the \texttt{RNS} code to compute the observables of rotating NSs.

\subsection{Rapidly rotating neutron star}

For a rotating NS, rotation can be treated as a perturbation when its effects on spacetime are not comparable to those of the star's self gravity~\citep{2001CQGra..18..969A}. However, this perturbation approach fails when the NS rotates rapidly. In this case, a two-dimensional, axially symmetric system must be solved to obtain the properties of rapidly rotating NSs.

We adopt the open-source \texttt{RNS} code~\citep{1989MNRAS.237..355K, 1994ApJ...422..227C, 1995ApJ...444..306S} to compute rapidly rotating NS models. The \texttt{RNS} code is widely used to solve the Einstein and hydrodynamic equations in an axially symmetric and stationary framework, given an input NS EoS~\citep{2013MNRAS.433.1903U, 2024ApJ...962...61M, 2025arXiv250818434S}. Once the central density and the axes ratio $r_p/r_e$, which is defined as the ratio between the polar coordinate radius $r_p$ and the equatorial radius $r_e$, are specified, the code solves the system and computes the corresponding observables. We list in Tab.~\ref{tab:observables} all observables output by \texttt{RNS} and used in the NN training. Note that our primary observables of interest are the mass, radius, and angular velocity; however, we also include other observables in the training in case they prove useful in the future work.

\begin{table}[ht]
    \centering
    \caption{List of observables output by the simulation code.}
    \label{tab:observables}
    \begin{tabularx}{\textwidth}{cccccc}
        \toprule
        \textbf{Symbol} & \textbf{Unit} & \textbf{Description} & \textbf{Symbol} & \textbf{Unit} & \textbf{Description} \\
        \midrule
        $M$ & $M_\odot$ & gravitational mass & $\phi_2$ & $10^{42}\,{\rm g\,cm^2}$ & quadrupole moment \\
        \addlinespace 
        $M_0$ & $M_\odot$ & baryonic mass & $h_+$ & km & height of co-rotating ISCO \\
        \addlinespace
        $R$ & km & circumferencial radius at equator & $h_-$ & km & height of counter-rotating ISCO \\
        \addlinespace
        $\Omega$ & $10^4\,{\rm s^{-1}}$ & angular velocity & $Z_p$ & -- & polar redshift\\
        \addlinespace
        $T/W$ & -- & rotational/gravitational energy & $Z_b$ & -- & backward equatorial redshift\\
        \addlinespace
        $J$ & $GM_\odot^2/C$ & angular momentum & $Z_f$ & -- & forward equatorial redshift\\
        \addlinespace
        $I$ & $10^{45}\,{\rm g\,cm^2}$ & moment of inertia & $\Omega_p$ & $10^4\,{\rm s^{-1}}$ & angular velocity of a particle at the equator\\
        \addlinespace
        $r_p/r_e$ & -- & axes ratio \\
        \bottomrule
    \end{tabularx}
\end{table}

\subsection{Causal neural network}

In this section, we introduce the structure of the neural networks to serve as a surrogate model. We separate our training data into three categories, \ie the static, the Keplerian, and the rotating configurations. Several observables in the static case are trivial and are therefore not included in our training data. In addition, the static and Keplerian configurations depend only on the EoS, whereas the rotating configuration also requires the axes ratio to be specified. These differences motivate training three separate models.

\begin{figure}
\vspace{-0.3cm}
{\centering
\includegraphics[width=0.99\textwidth]{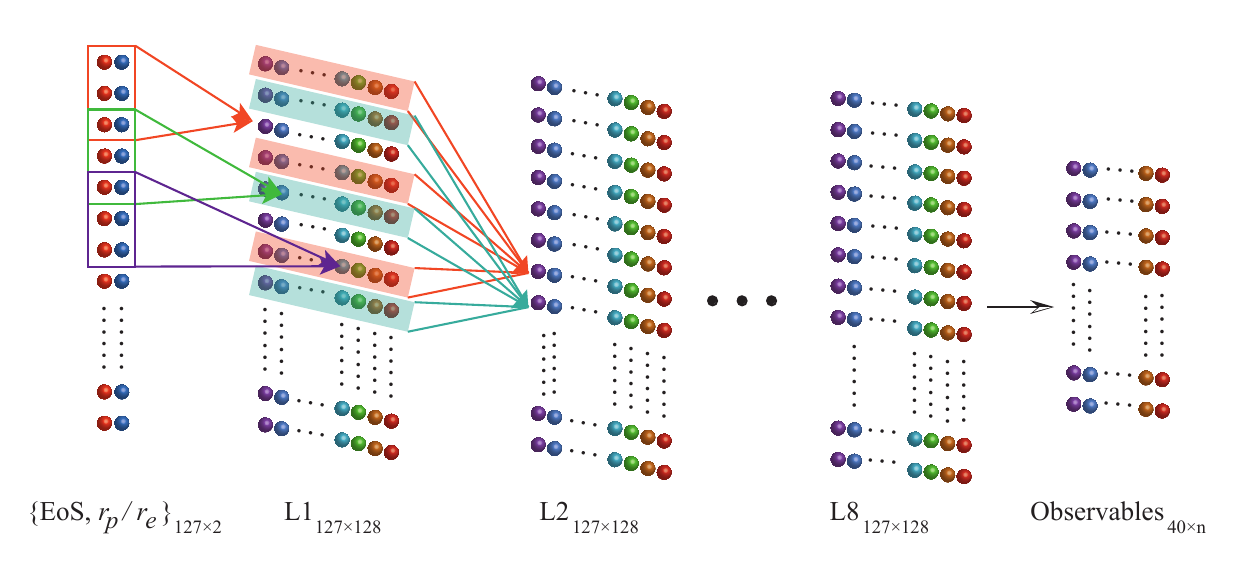}}
\caption{The structure of causal convolutional networks for rotating configurations. For the other two networks (static and Keplerian), the structures are slightly different: they take only the one-array EoS input and use fewer (5) hidden layers and fewer (64) channels. Note that the different colored arrows in latent layes illustrate the causal receptive field of individual nodes at different sequence position.
}\label{fig:nn_rot}
\vspace{-0.1cm}
\end{figure}

In Fig.~\ref{fig:nn_rot}, we show the neural network model for rotating configurations. The EoS dependence of NS properties is analogous to that of causal time-series data: the properties of a NS at a given central density depend only on the EoS at densities up to that central density, not on higher-density behavior. Therefore, we adopt the causal convolutional neural network~\citep{2016arXiv160903499V, 2022JCAP...08..071S, 2023PhRvD.107h3028S}, which retains this chronological-like feature of the system, to learn the \texttt{RNS} solver. The cold equilibrated one-dimensional EoS has only one degree of freedom: once the pressure as a function of baryon number density $p(n_{B})$ is known, other variables can be determined through thermodynamical relations. we therefore use an array with 127 elements to represent the EoS input, corresponding to the pressure at discrete $n_{B}$ value, with $n_{B}$ logarithmically spaced from $2.572\times 10^{-4},{\rm fm^{-3}}$ to $1.6,{\rm fm^{-3}}$. For the rotating case, the additional $r_p/r_e$ input is expanded to a constant array with the same length as the EoS pressure array, and the two together form a two-dimensional input array. Note that our input EoS includes the inner crust part, while the outer-crust EoS is assumed to be the BPS EoS.

As shown in Fig.~\ref{fig:nn_rot}, each new layer is obtained by convolving only the preceding part of the previous layer (\ie using only pressure information corresponding to lower densities). This structure ensures that the output at a given position depends only on earlier elements of the input sequence, and thus preserves the causality of the sequence. In addition, we introduce dilation to expand the receptive field. Dilated convolution enables the kernel to sample the input sequence at intervals with stride $d$, thereby exponentially expanding the dependent domain without increasing the number of parameters. \eg consider a one-dimensional EoS input $\bm{x}$, the output $\bm{y}$ after a dilated causal convolution can be expressed as
\begin{eqnarray}
  \label{eq:nn_cc}
  y = \sum_{i=0}^{k-1} \bm{W}_i \cdot \bm{x}_{t-i*d} + \bm{b},
\end{eqnarray}
where $\bm{W}$ and $\bm{b}$ denote the convolutional kernal weights and bias, and $k$ is the kernal length. By iterating the convolution in Eq.~(\ref{eq:nn_cc}) and progressively increasing the dilation rate layer by layer, such that dilation $d=k^{l-1}$ for the $l$-th layer, the network can incorporate all lower-density data while maintaining computational efficiency.

We construct three networks for the static, Keplerian, and rotating configurations~\footnote{The NN reconstruction developed in this work is publicly available at \url{https://github.com/zzhu-astro/NN_RNS}.}.
Figure~\ref{fig:nn_rot} shows the architecture of the rotating case. The static and Keplerian networks share the main components of the rotating networks. However, for these two cases, the axes ratio does not need to be included as an input. Additionally, the amount of training data is smaller, and a more compact network is sufficient to achieve accurate reconstruction. We therefore reduce the number of layers and reduce the number of channels in each layer. We have tested different sizes of networks and found that this configuration provides the best balance between efficiency and accuracy.

The model is implemented in the \textit{PyTorch} framework and employs the mean squared error (MSE) as the loss function to quantify the discrepancy between the predicted values and the ground truth. The overall architecture consists of a stack of 8 (rotating) or 5 (static and Keplerian) dilated causal convolutional blocks, each comprising a dilated causal convolution layer followed by batch normalization and the ReLU activation function. The first convolutional layer uses an input-channel number equal to the width of the input array (2 for the rotating network and 1 for the other two networks), and the number of channels is increased to 128 (rotating) or 64 (static and Keplerian) in the subsequent layers. The output layer, whose number of output channels matches the number of observables, is applied without batch normalization or an activation function. Finally, we got 40 sets of observables corresponding to models whose central pressure is given by the values in the input EoS array.

\section{Network Training}
\label{sec:training}

The original data produced by the \texttt{RNS} code includes a large amount of unreliable outputs. In some cases, the code may fail to converge after a certain number of iterations when solving the 2D axially symmetric equations and therefore yield incorrect solutions. The solver may also converge to unphysical solutions of the equations, which is also undesirable. Fortunately, most of these unreliable data occur for low-mass NSs or along the unstable branch (which has a higher central density than that of the maximum-mass NSs). We can remove most of the unreliable output by eliminating data with excessively large radii and masses. Specifically, we set the upper limit of the mass as $3.5M_\odot$ ($5.0M_\odot$) and of the radius to $50$km ($80$km) for the cases of static (Keplerian-rotating) NSs. However, these upper limits are typically EoS-dependent, and adopting a universal limit for all EoS can miss certain amount of unreliable data. Therefore, we further remove all the unconverged data, which covers most of the remaining unreliable output and yields final trained networks with sufficiently low errors.

We use the \texttt{RNS} code to solve rotating NS systems for $20{,}000$ EoS and obtain the corresponding observables, which are then split into training ($80\%$) and testing ($20\%$) sets to train these networks. We use the mean squared error (MSE) loss function and Adam optimizer with a learning rate of $10^{-4}$, and run the training process on Intel Xeon Platinum 8358 CPUs for more than 2000 epochs. The training process stops when the losses converge, and the final losses for three networks on the training (testing) datasets are: static network $4.0\times 10^{-4}$ ($3.0\times 10^{-4}$), Keplerian network $3.5\times 10^{-4}$ ($2.5\times 10^{-4}$) and rotational network $9.5\times 10^{-5}$ ($6.0\times 10^{-5}$) for training (testing) dataset.

\begin{figure}
\vspace{-0.3cm}
{\centering
\includegraphics[width=0.333\textwidth]{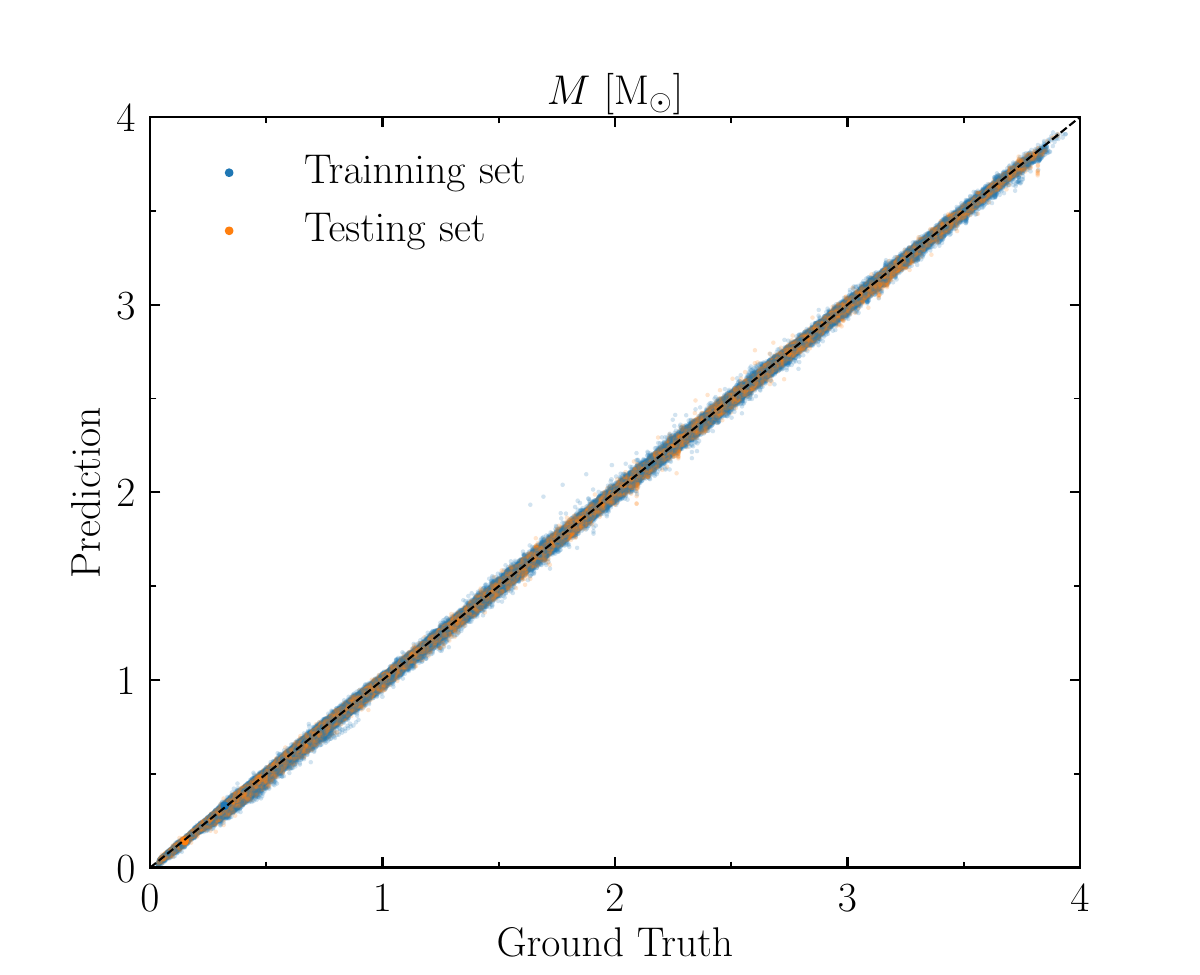}
\includegraphics[width=0.333\textwidth]{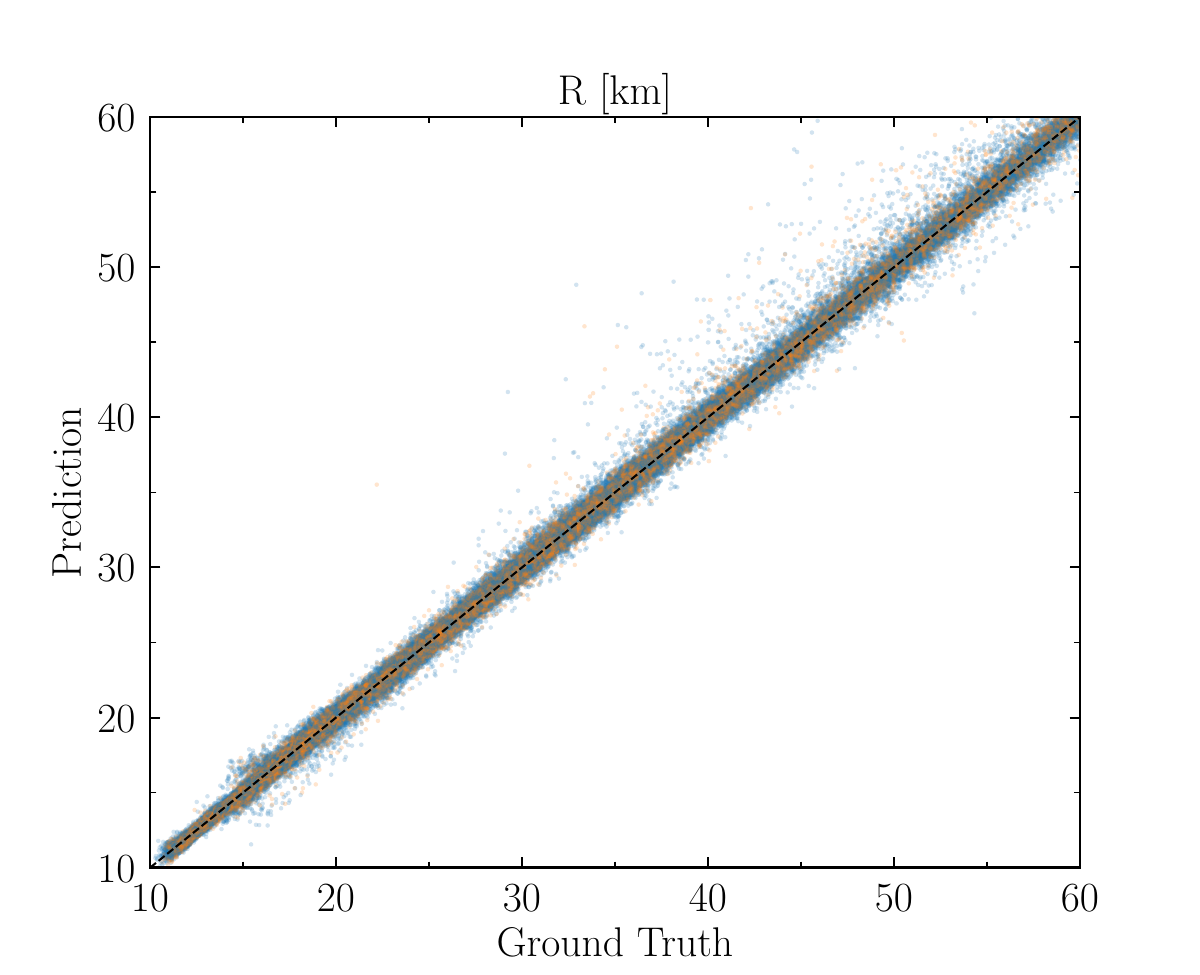}
\includegraphics[width=0.333\textwidth]{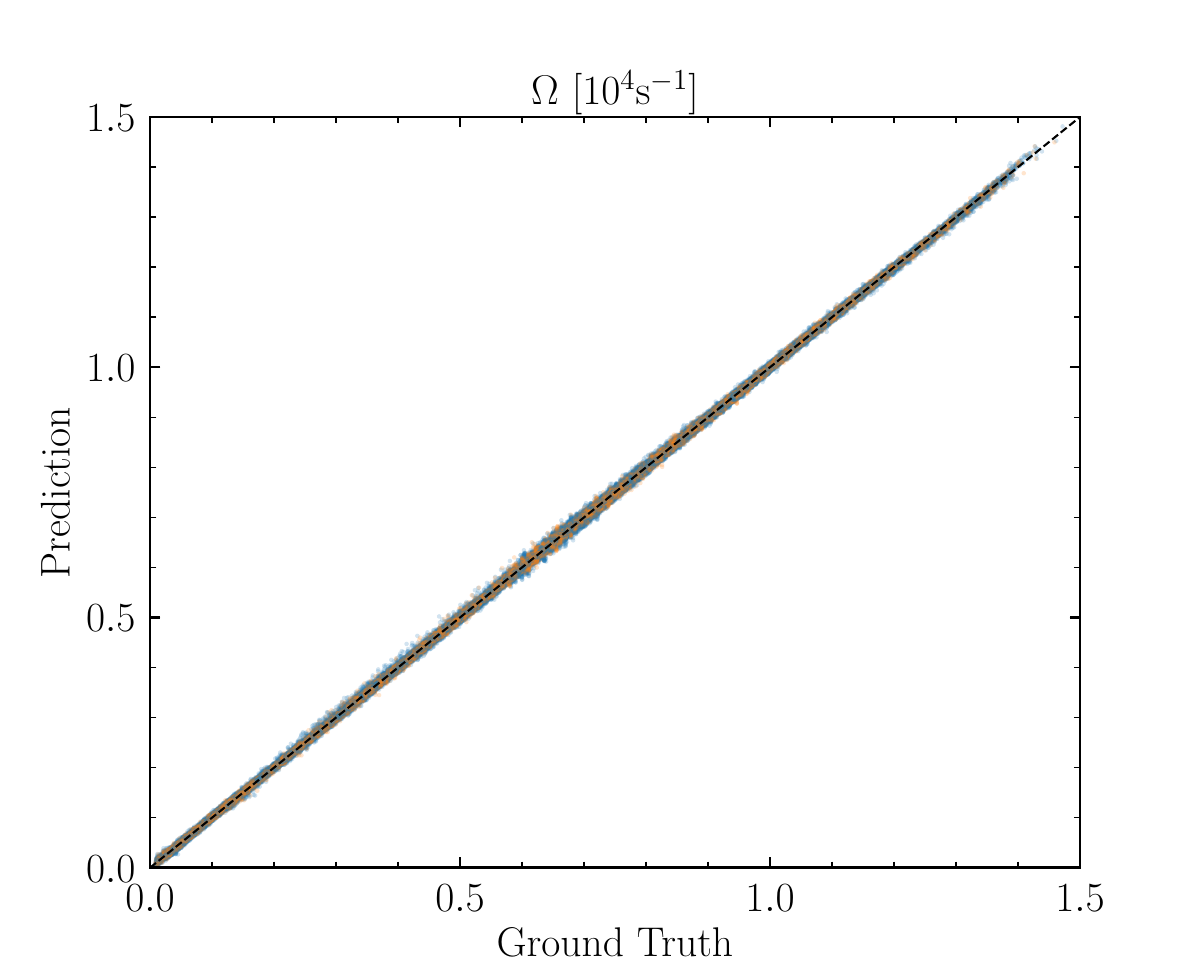}}
\caption{The parity plots for the Keplerian-rotation network compare the ground truth and the predictions, shown on the horizontal and vertical axes, respectively. The data points in the training and test sets are denoted by blue and orange dots. The identity lines are shown as black dashed lines. The results for gravitational mass, equatorial radius and angular velocity are shown in the left, middle and right panels, respectively.}\label{fig:kep_pt}
\vspace{-0.1cm}
\end{figure}

As an example of the static and Keplerian cases (which have almost the same structure of networks), we display the parity plots in Fig.~\ref{fig:kep_pt}. This figure compares the ground truth values and predictions for the gravitational mass, circumferential radius, and angular velocity of the NSs in the left, middle and right panels, respectively. Data points in training and testing sets are indicated by blue and orange dots. The identity lines are shown as black dashed lines. The mean relative errors in the reconstruction of mass, radius and angular velocity are less than $0.6\%$, $0.8\%$ and $0.4\%$, respectively, for NSs with $M\gtrsim M_{\odot}$ in both training and testing datasets.

We find that most of the dots in the left and right panels are well aligned with the identity line, while relatively larger deviations are observed in the middle panel for the radius reconstructions. More specifically, the deviation becomes significant when the radius is larger than $\sim 15$km. NSs with radii larger than $15$km typically correspond to small masses ($\lesssim 1 M_\odot$). These low-mass NSs are not our primary concern and only account for a small fraction of the full dataset. In addition, other observables, including the mass and angular velocity for these large-radius NSs, are relatively small, so their deviations are less apparent in the bottom-left corners of the corresponding panels. In general, the good alignment of data points for typical NSs ($\gtrsim 1 M_\odot$) with identity lines indicates that the Keplerian-rotating network has been trained to a sufficient accuracy.

\begin{figure}
\vspace{-0.3cm}
{\centering
\includegraphics[width=0.333\textwidth]{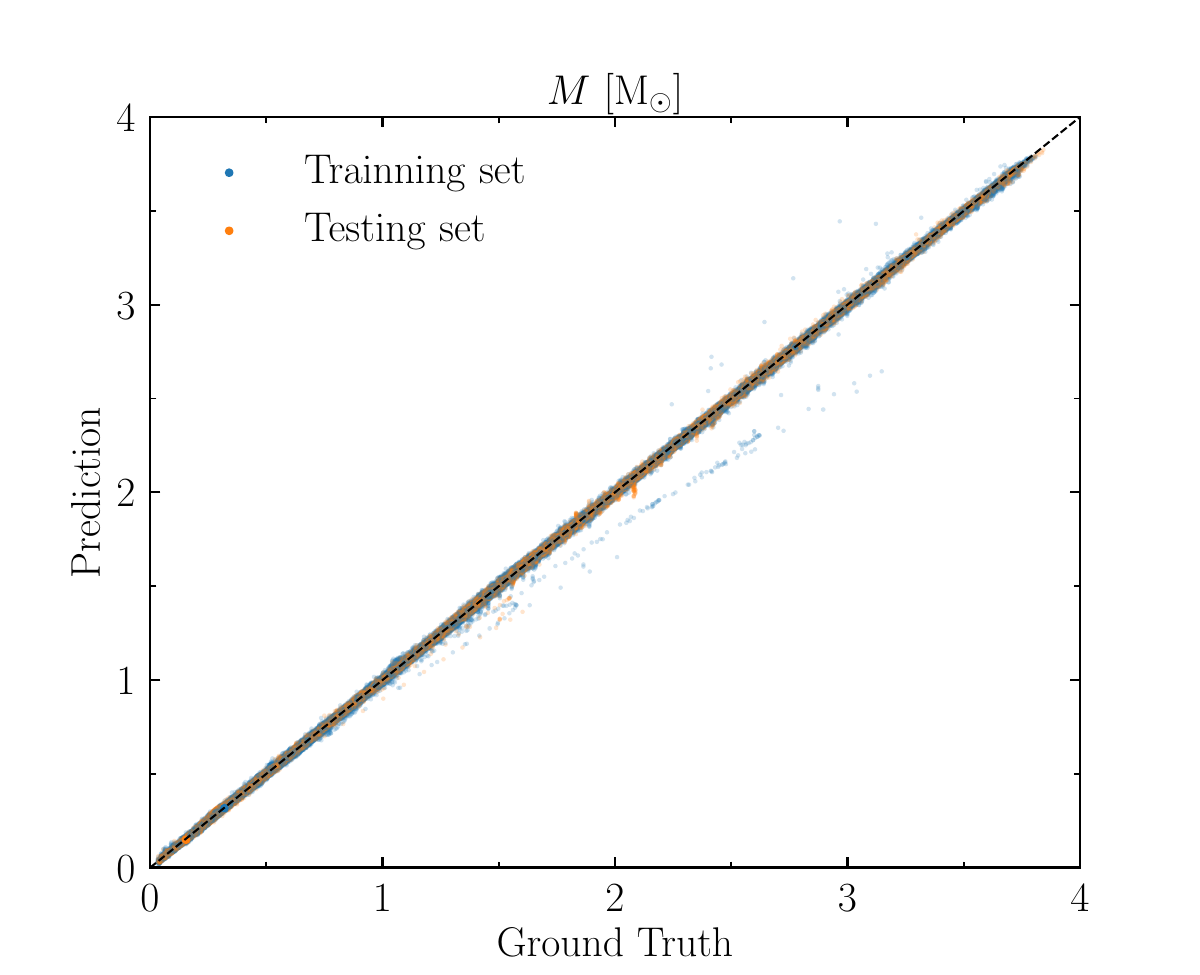}
\includegraphics[width=0.333\textwidth]{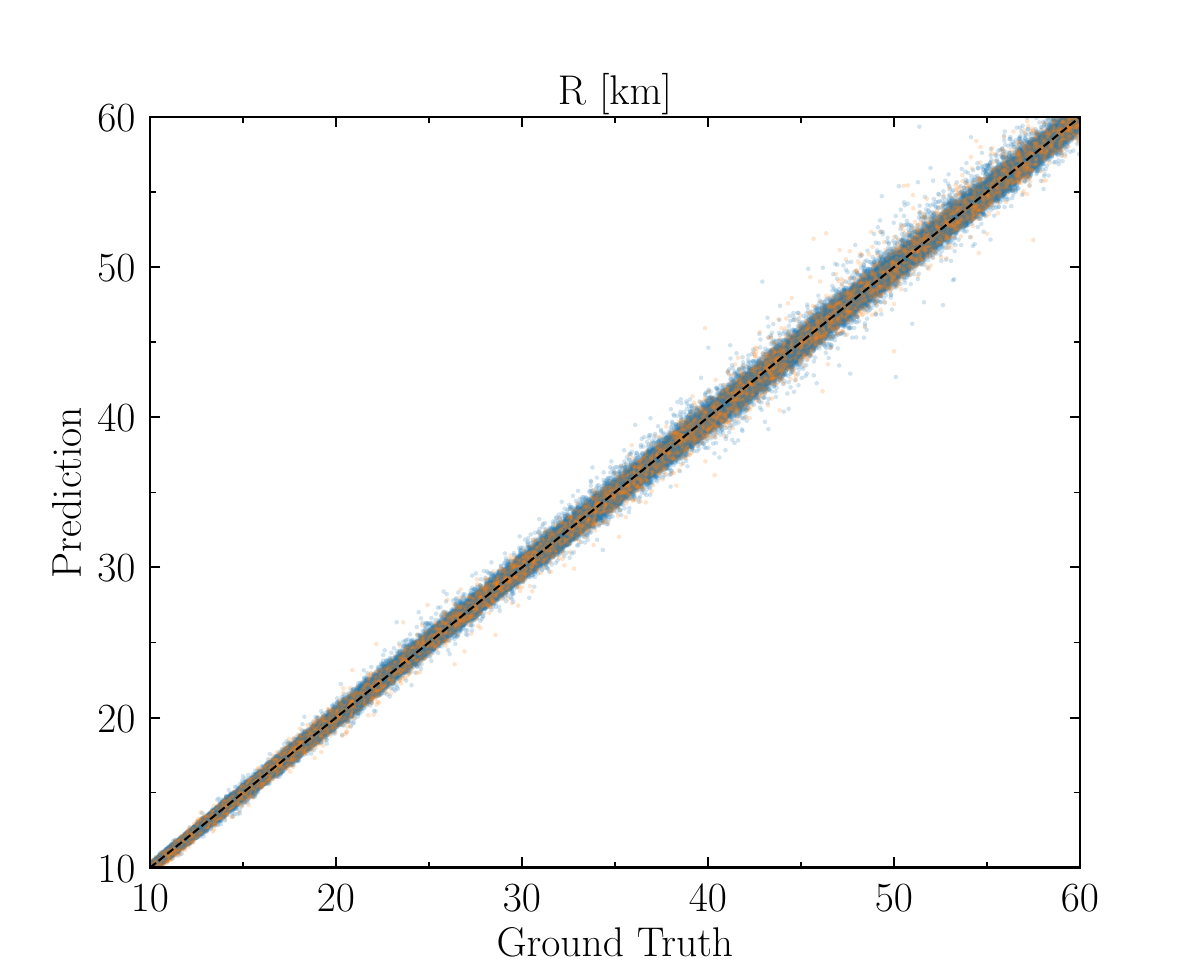}
\includegraphics[width=0.333\textwidth]{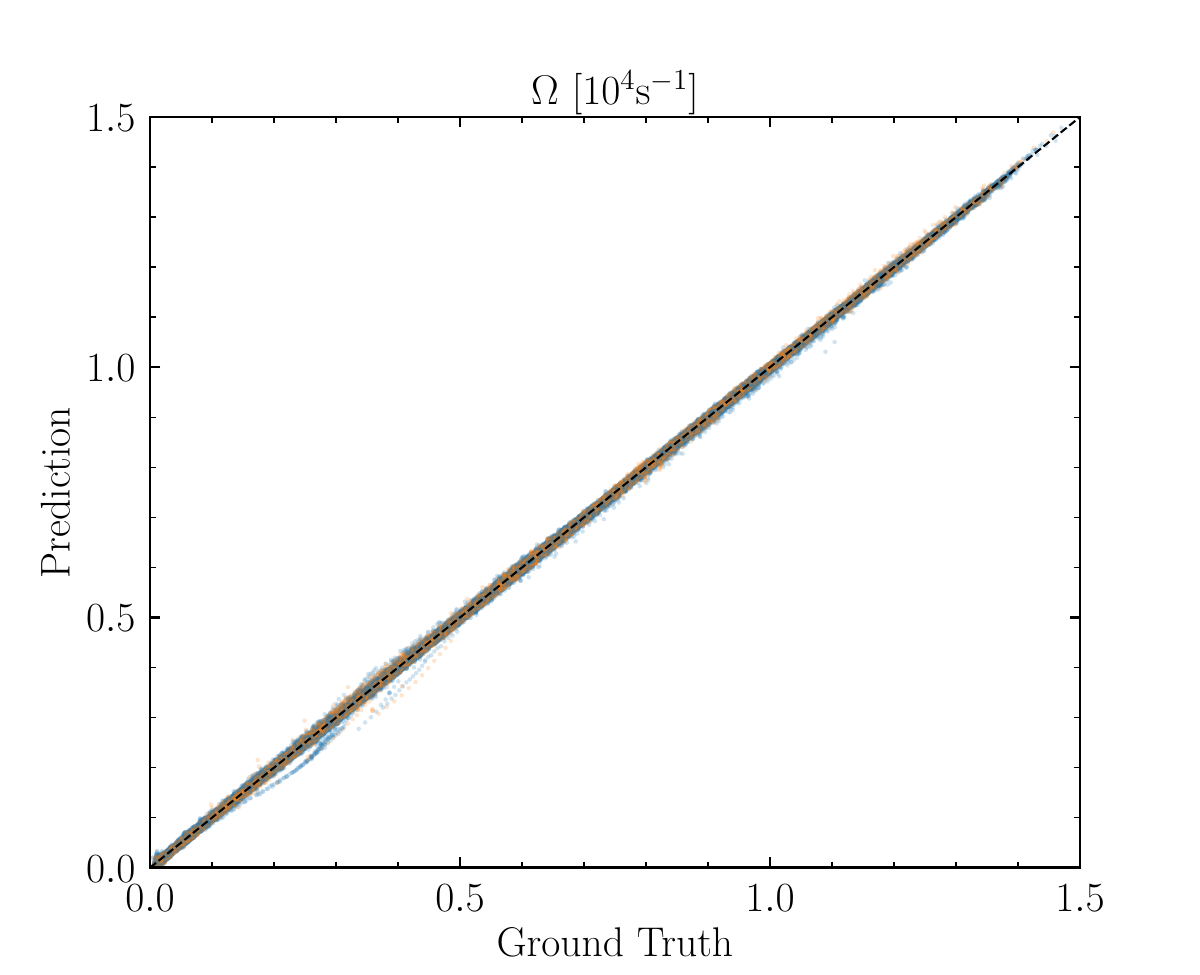}}
\caption{The same parity plots as in Fig.~\ref{fig:kep_pt}, but for the rotating network.}\label{fig:rot_pt}
\vspace{-0.1cm}
\end{figure}

However, the situation for rotating stars can be more complicated. Removing unconverged data and data points with excessively masses and radii cannot fully eliminate the contamination from unphysical results. We therefore impose additional constraints on the mass, radius, and angular velocity of rotating NSs, \eg, requiring the monotonicity of the angular velocity $\Omega$ with the axes ratio $r_p/r_e$. After thoroughly cleaning the original rotating dataset, we train the rotating network using the remaining high-quality, physically reliable data until sufficiently low errors are achieved. In Fig.~\ref{fig:rot_pt}, we present the training results of the rotating network in this parity plot, which share the same labels as in Fig.~\ref{fig:kep_pt}.

Thanks to the thorough data cleaning of the rotating-NS dataset, we find that both the training and testing data points from the rotating network are more tightly aligned with identity lines than those from the Keplerian network. However, a small number of data points in the testing set remain significantly offset from the line, as shown in the left panel. These points have been double-checked and confirmed to be unphysical results. Nevertheless, the network has been trained to sufficient accuracy to reconstruct the properties of rotating NSs. Therefore, we did not remove them and redo the training process.

\section{Results and discussions}
\label{sec:results}
The primary observables of interest are the mass, radius, and angular velocity of NSs, as these observables can be directly measured by telescopes or detectors. However, other observables may also be important for some cases. We present and discuss the results of these additional observables in the Appendix~\ref{sec:appendix1}.

Note that our training data for the rotating network are available only at 10 discrete values of axes ratio $r_p/r_e$ (from $0.5$ to $0.95$ in steps of $0.05$). For a given EoS, we therefore first evaluate the rotating network predictions at these ten baseline axes ratios. We then combine these results with the static and Kepler-rotating models and use interpolation to obtain the observables at intermediate rotation states (i.e., other values of $r_p/r_e$).

\begin{figure}
\vspace{-0.3cm}
{\centering
\includegraphics[width=0.302\textwidth]{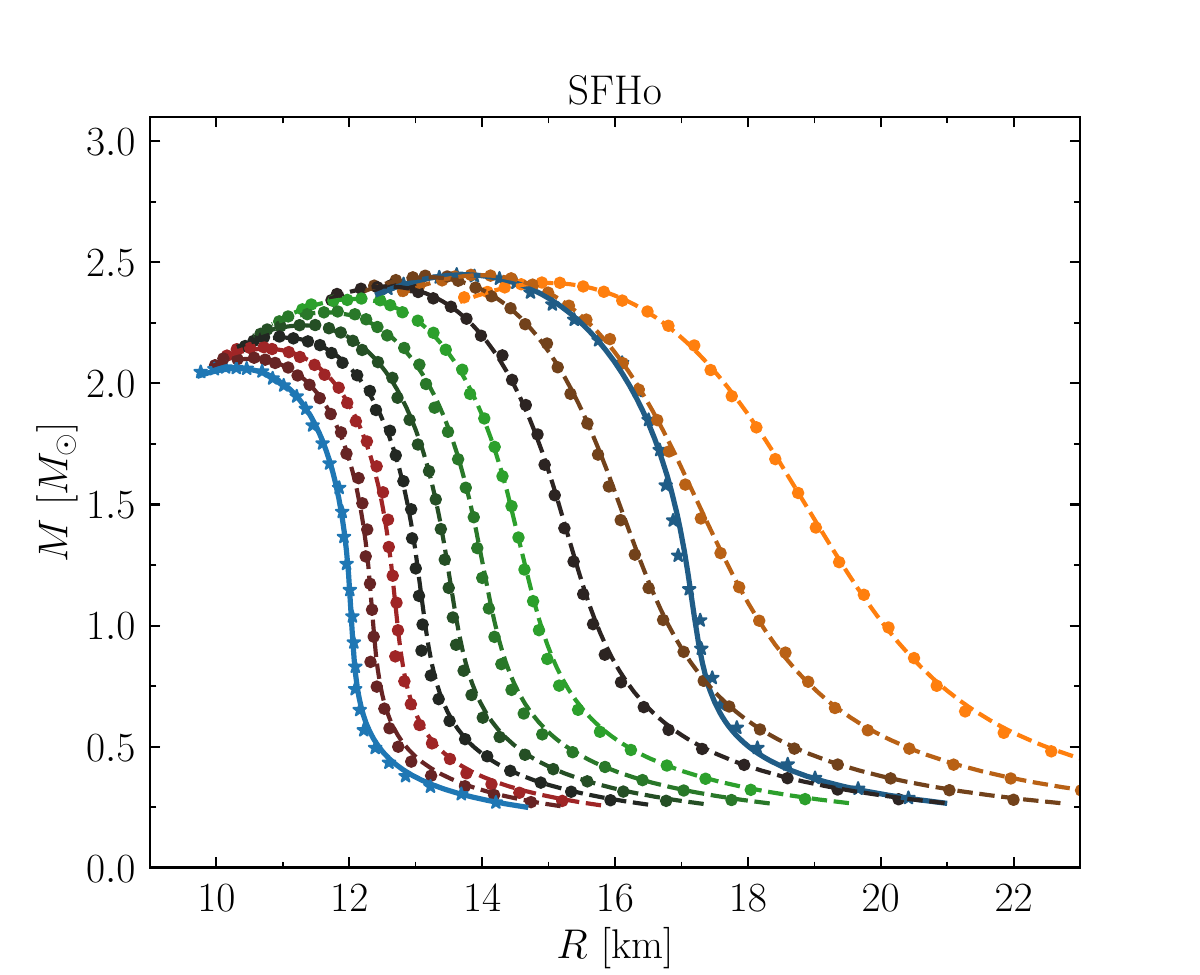}
\includegraphics[width=0.302\textwidth]{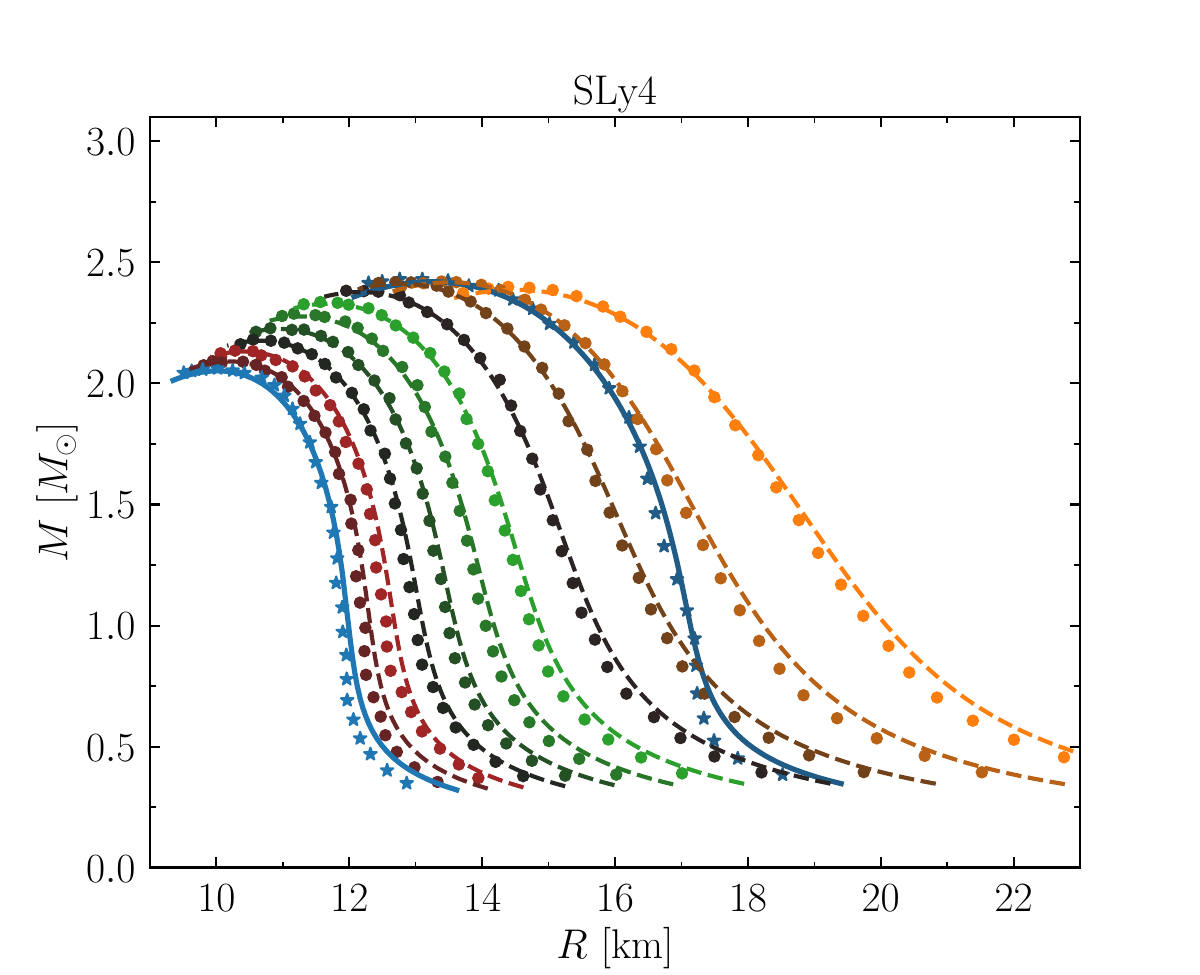}
\includegraphics[width=0.396\textwidth]{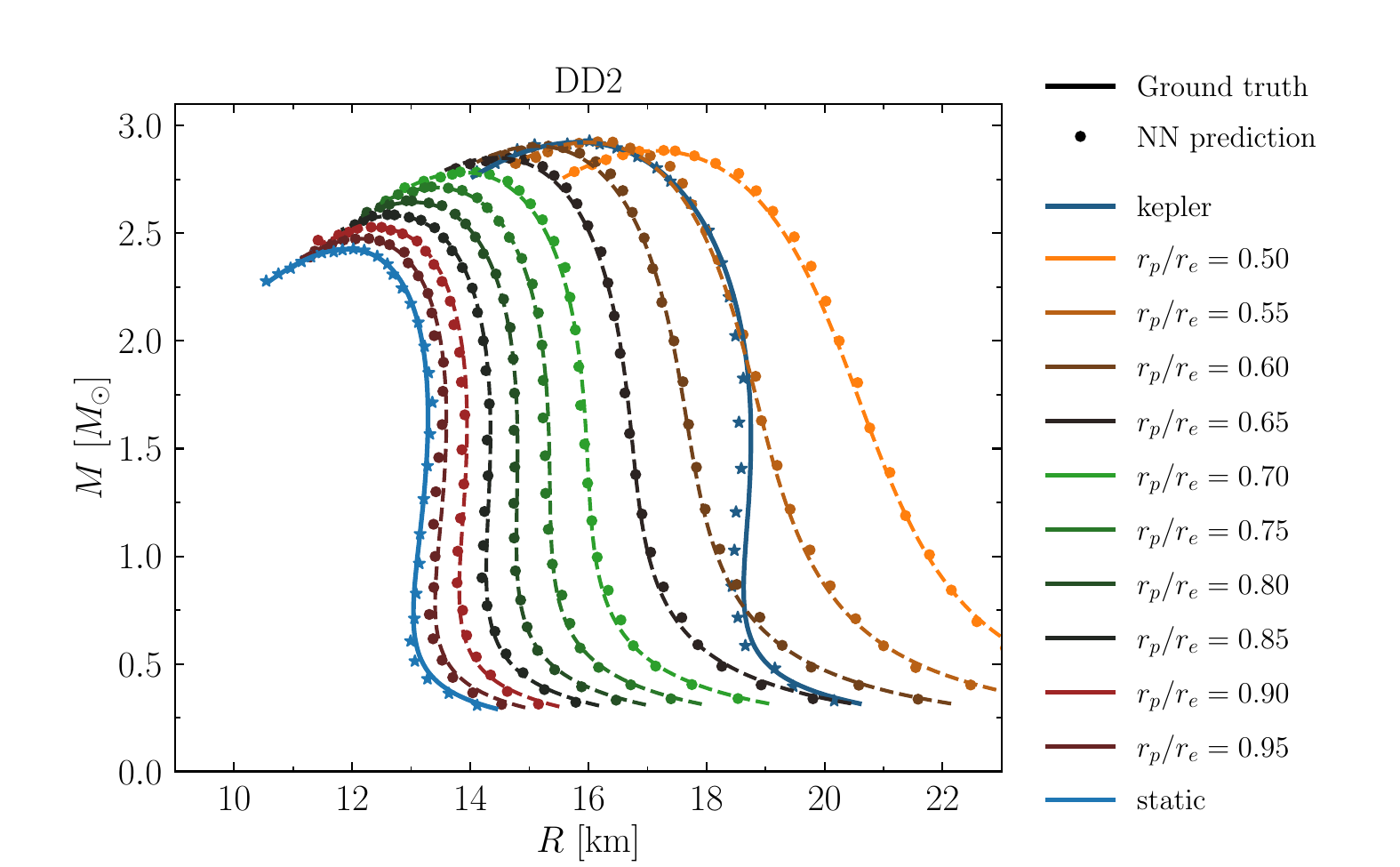}}
\caption{The mass-radius relations of rotating NSs evaluated at the baseline axes ratios for SFHo (left), SLy4 (middle), and DD2 (right) EoS. The \texttt{RNS} ground truth and NN predictions are shown as lines and dots, respectively. Colors indicate different axes ratios, as well as the static and Kepler cases.}
\label{fig:r_ratio_mr}
\vspace{-0.1cm}
\end{figure}

In Fig.~\ref{fig:r_ratio_mr}, we compare the mass-radius relations from the \texttt{RNS} ground truth (lines) and NN predictions (dots) for the static model, the Kepler model, and the rotating models evaluated at the baseline axes ratios. The results for three representative EoS (SFHo~\citep{2013ApJ...774...17S}, SLy4~\citep{1998NuPhA.635..231C}, and DD2~\citep{2010PhRvC..81a5803T}) are shown in the left, middle, and right panels, respectively. Overall, the NN predictions agree well with the ground truth, and provide a substantial gain in computational efficiency. In general, evaluating one EoS with the NN takes $\sim 50$,ms, which is a significant acceleration compared to the \texttt{RNS} computations that take $\sim 30$,min. The SLy4 results show relatively lower precision than the other EoSs, mainly due to the EoS input: the \texttt{RNS} outputs for the SLy4 EoS contain lots of invalid and unphysical data points, making SLy4 a representative example of the deviating points seen in Fig~.\ref{fig:rot_pt}. We also note that for small $r_p/r_e$, the radius can exceed the Keplerian-rotation limit. Although such configurations are physically inaccessible, they remain numerically valid and are useful for interpolating intermediate rotation states. We therefore retain them in the training data and in the corresponding results.

\begin{figure}
\vspace{-0.3cm}
{\centering
\includegraphics[width=0.333\textwidth]{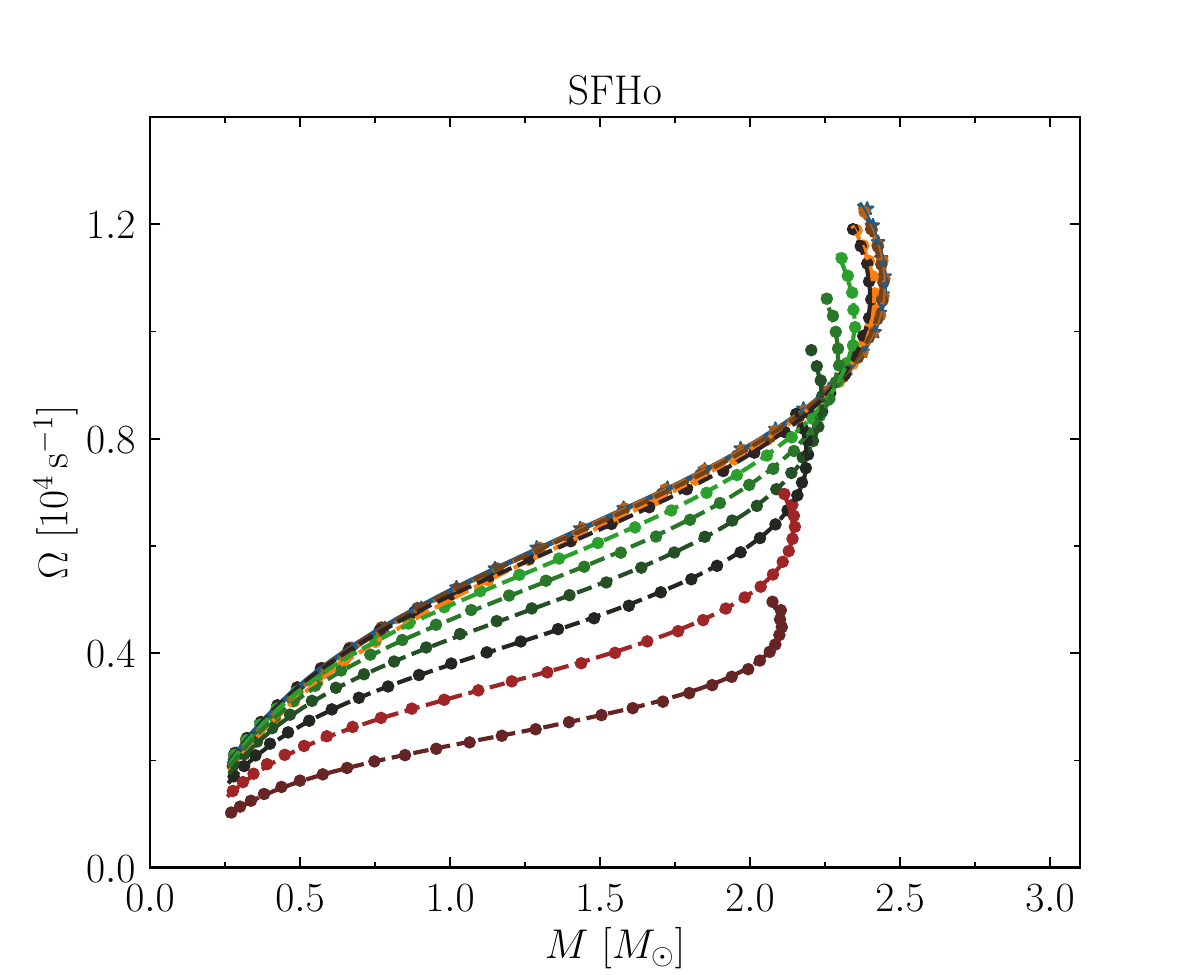}
\includegraphics[width=0.333\textwidth]{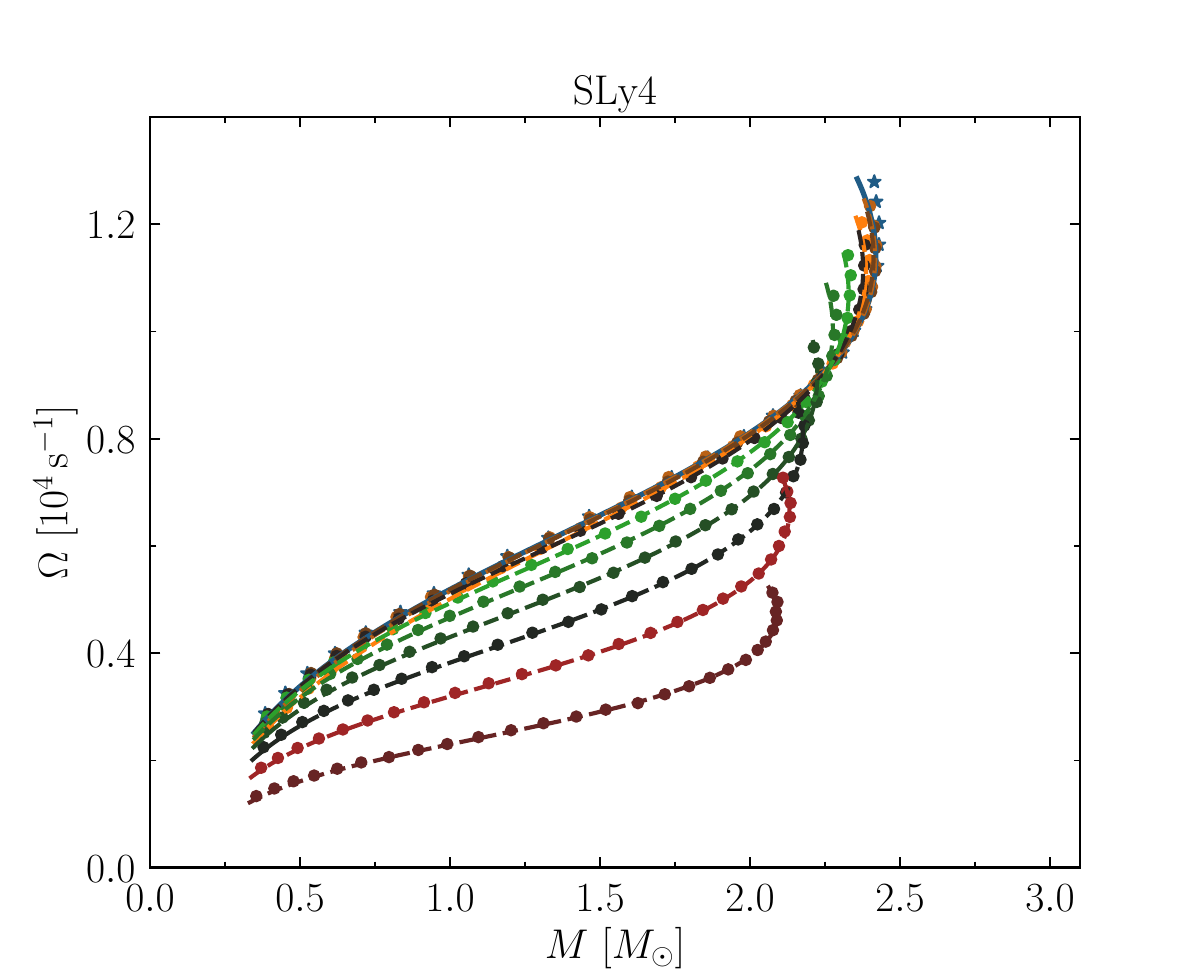}
\includegraphics[width=0.333\textwidth]{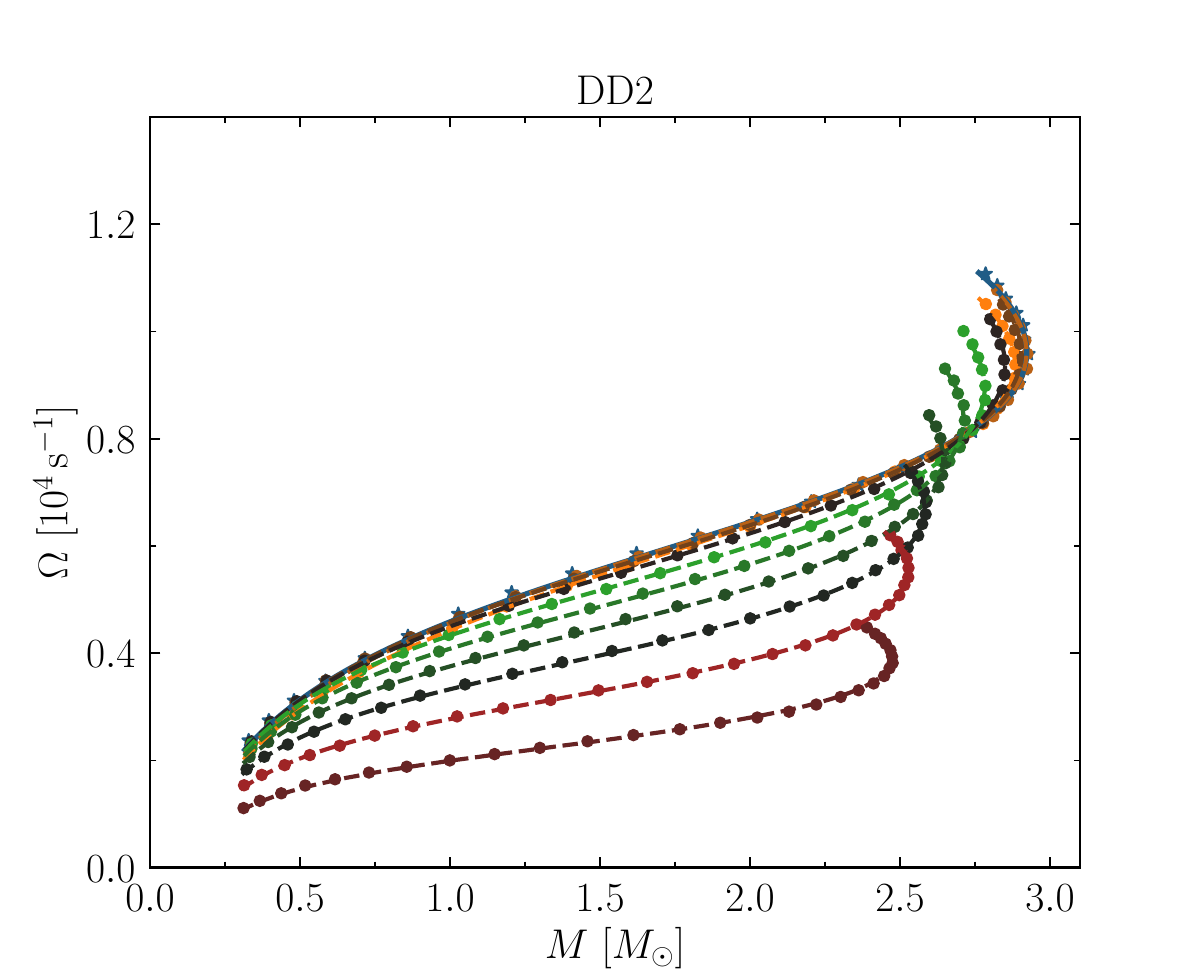}}
\caption{The angular velocity as a function of mass for different EoS. The labels are the same as in the previous figures.
}\label{fig:r_ratio_Omega}
\vspace{-0.1cm}
\end{figure}

Another important observable is the NS angular velocity, which can be directly measured through the pulsar timing. In Fig.\ref{fig:r_ratio_Omega}, we show the angular velocity $\Omega$ as a function of mass for three EoSs. The labels in this figure are the same as in Fig.\ref{fig:r_ratio_mr}. We find that the prediction of angular velocity is more accurate than that of the radius, consistent with the results shown in Fig.\ref{fig:rot_pt} and Fig.\ref{fig:kep_pt}. In particular, the $\Omega$ values in the training dataset are more compactly and evenly distributed (from $\sim 0$ to $\sim 1.5$) than the radii, which span a wide range from $\sim 9$km to $\sim 60$km, with most valid data concentrated at radii below $\sim 25,\mathrm{km}$. Although the observables are normalized before training, the presence of very large radii, which correspond to low-mass NSs, can degrade the reconstruction accuracy for the smaller radii of typical NSs, due to their large numerical values and the lower reliability of the computations for low-mass configurations.

Note that the $\omega$-$M$ relation does not include the static model due to its vanishing and trivial values of angular velocity.
In contrast to the radius, we find that for the stars with axes ratios exceeding Kepler limit, their angular velocity does not exceed the Keplerian value by much. As a result, the $\Omega$-$M$ lines converge toward the Keplerian lines as $r_p/r_e$ increase. This behavior follows directly from the definition of Keplerian rotation, which corresponds to the limiting angular velocity at which the centrifugal force balances gravity. Therefore, even a small excess of $\Omega$ over the Keplerian angular velocity can lead to a large change in $r_p/r_e$. This is also the reason why we use $r_p/r_e$, rather than $\Omega$, as an input variable in our training data, interpolation in terms of $\Omega$ would be highly inaccurate near the Keplerian limit.

\begin{figure}
\vspace{-0.3cm}
{\centering
\includegraphics[width=0.333\textwidth]{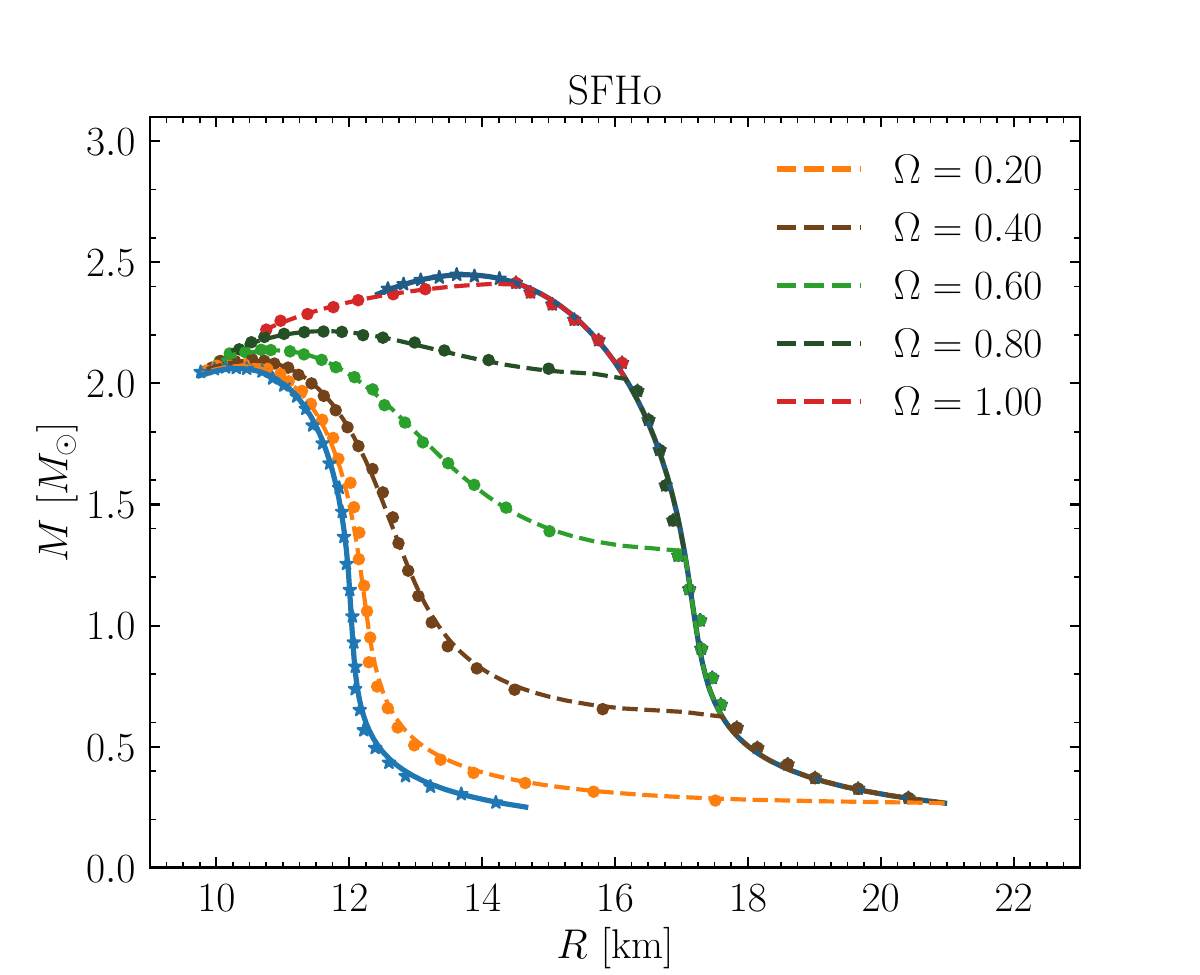}
\includegraphics[width=0.333\textwidth]{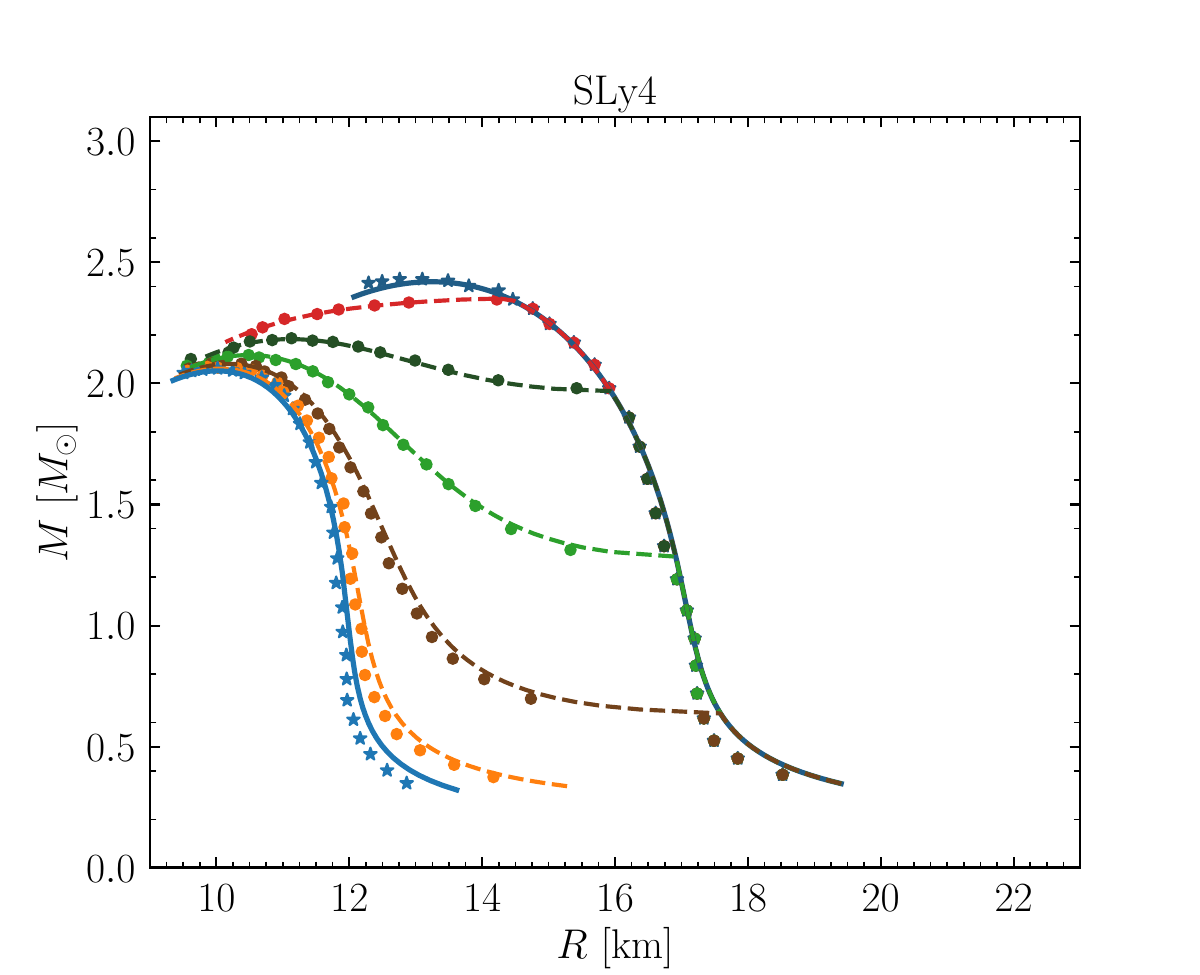}
\includegraphics[width=0.333\textwidth]{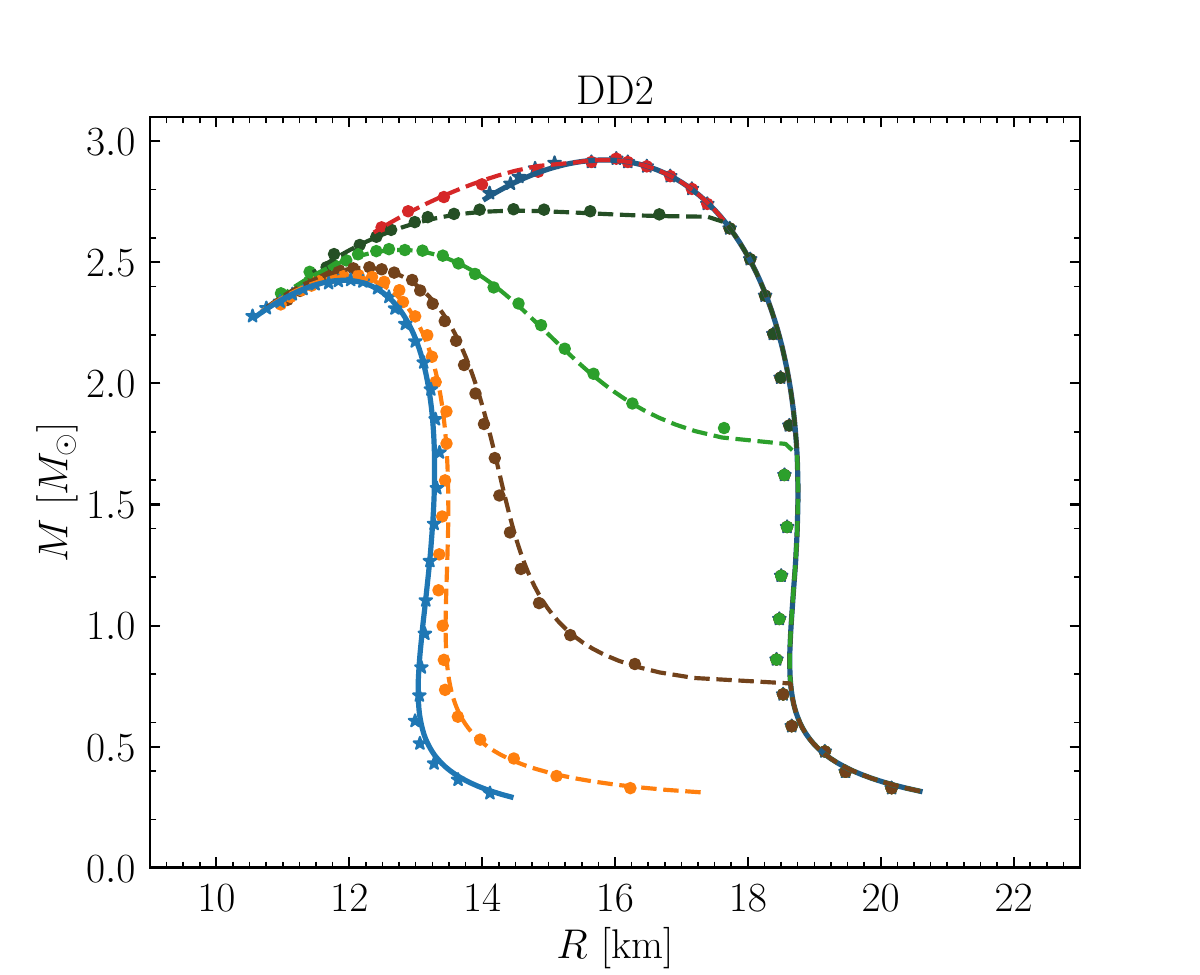}}
\caption{The mass-radius relations for configurations with five values of $\Omega$ ($0.2$, $0.4$, $0.6$, $0.8$, $1.0$, in units of $10^4{\rm s^{-1}}$). The static and Kepler models are also included, and the labels for them, as well as for the \texttt{RNS} ground truth and the NN predictions, are the same as in previous figures.}\label{fig:Omega}
\vspace{-0.1cm}
\end{figure}

However, in many applications, especially for the EoS inference where the masses of rotating pulsars serve as an upper bound on the NS maximum mass, we would like to compute the stellar properties at a specific angular velocity. This requires interpolating in $\Omega$ using the results at discrete axes ratios in our NN models. To test the accuracy of this procedure, we show the mass-radius relations of NSs at several fixed values of $\Omega$ in Fig.~\ref{fig:Omega}. The labels for \texttt{RNS} ground truth, the NN predictions, and the static and Kepler models are the same as in previous figures. We display $M$-$R$ for configurations with five values of $\Omega$ ($0.2$, $0.4$, $0.6$, $0.8$, $1.0$, in units of $10^4{\rm s^{-1}}$). Note that each $\Omega$ line ends on the Keplerian line, because we enforce $\Omega$ to be smaller than the Keplerian angular velocity during interpolation. 

We see that the accuracy is mildly lower than that in Fig.~\ref{fig:r_ratio_mr} due to the interpolation. However, if we focus on the lines near the maximum mass, which is our primary concern since this quantity is directly used in EoS inferences, the errors of NN prediction remain at a high level of precision. This conclusion also holds for the SLy4 EoS. Therefore, our interpolation procedure provides sufficiently accurate predictions in the region of interest.

\section{Conclusion}
\label{sec:conclusion}

Rotation is ubiquitous in neutron stars. For rapidly rotating systems, such as millisecond pulsars, rotation can significantly affect NS properties. However, an accurate and self-consistent computation of rotating NSs requires solving a two-dimensional axially symmetric system and can take several minutes per model. Because of this computational cost, such calculations cannot be directly used in inference analyses that involve rapidly rotating NS sources.

Therefore, we construct causal convolutional neural networks, which preserve the chronological-like dependence of NS properties on the EoS, to reconstruct the observables of fast-rotating NSs from the EoS. We construct and train three networks to represent the static, Keplerian and rotating models of NSs. We use the \texttt{RNS} code to solve rotating NS systems for $20,000$ EoS and obtain the corresponding observables, which are then split into training ($80\%$) and testing ($20\%$) sets to train these networks.

We validate our networks using three representative EoSs (SFHo, SLy4, and DD2) by comparing the mass--radius and mass--angular-velocity relations from \texttt{RNS} with 
the corresponding NN predictions for the static, Keplerian, and rotating configurations evaluated at 10 discrete values of the axes ratio $r_p/r_e$. The NN reproduces the 
\texttt{RNS} results with good accuracy across all three set of EoSs. We further validate the interpolation procedure used to obtain configurations at fixed angular velocity from the discrete-$r_p/r_e$ predictions: although the accuracy is mildly reduced relative to the direct $r_p/r_e$ evaluation, it remains sufficiently high in the mass range of 
primary interest for EoS inference. The computational efficiency of the NN is also remarkable: evaluating all NS configurations for a single EoS takes only ${\sim}50\,$ms, 
compared to ${\sim}30\,$min for the \texttt{RNS} solver --- a speedup of more than three orders of magnitude. These results demonstrate that our NN provides a fast and accurate 
emulator for rapidly rotating NS calculations, and is well-suited for large-scale inference analyses, particularly when combined with automatic differentiation frameworks~\citep{2025PhRvD.111g4026L,2025PhRvD.112d3037W}.

\section*{Acknowledgements}

We thank Richard O'Shaughnessy, Kaiming Cui, Ming-zhe Han, Shuzhe Shi for useful discussions and comments.
We thank the DEEP-IN working group at RIKEN-iTHEMS for support in the preparation of this paper.
This research is supported by the Start-Up Fund for new Ph.D. Researchers of Suzhou Chien-Shiung Institute of Technology, the National Natural Science Foundation of China (grant No. 12203033), the China Postdoctoral Science Foundation (No. 2022M712086 and BX20220207). L.W. is supported by the RIKEN-TRIP initiative (RIKEN-Quantum), JSPS KAKENHI Grant No. 25H01560, and JST-BOOST Grant No. JPMJBY24H9.

We acknowledge the open-source \texttt{RNS} code, originally developed by Nikolaos Stergioulas and collaborators, for constructing rapidly rotating neutron star models. The code is publicly available at \url{https://github.com/cgca/rns}. The \texttt{RNS} simulations and neural network training in this work were carried out on Siyuan Mark-I, supported by the Center for High Performance Computing at Shanghai Jiao Tong University.


\bibliographystyle{aasjournal}
\bibliography{rns_nn}

\appendix
\section{other observables}
\label{sec:appendix1}

\begin{figure}
\vspace{-0.3cm}
{\centering
\includegraphics[width=0.333\textwidth]{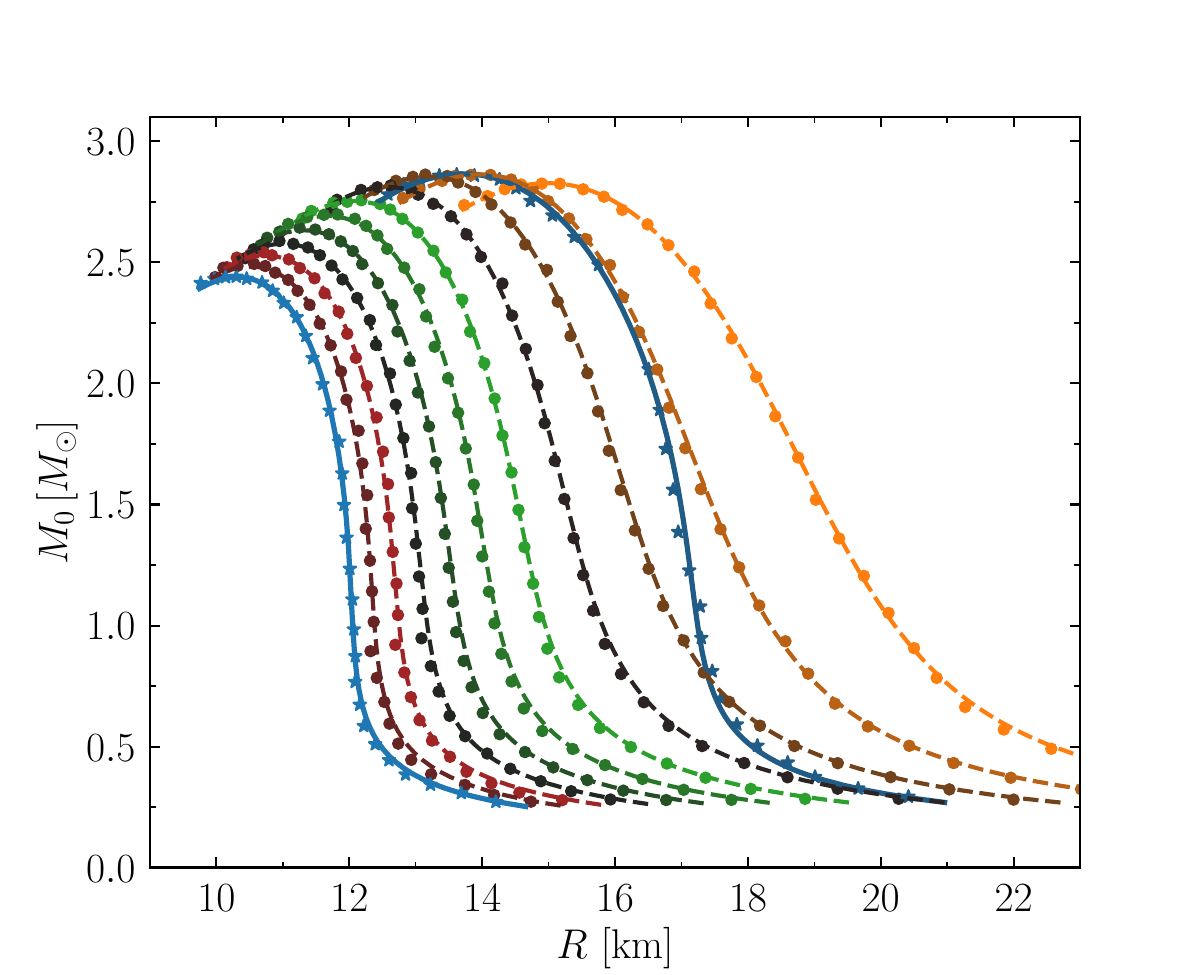}
\includegraphics[width=0.333\textwidth]{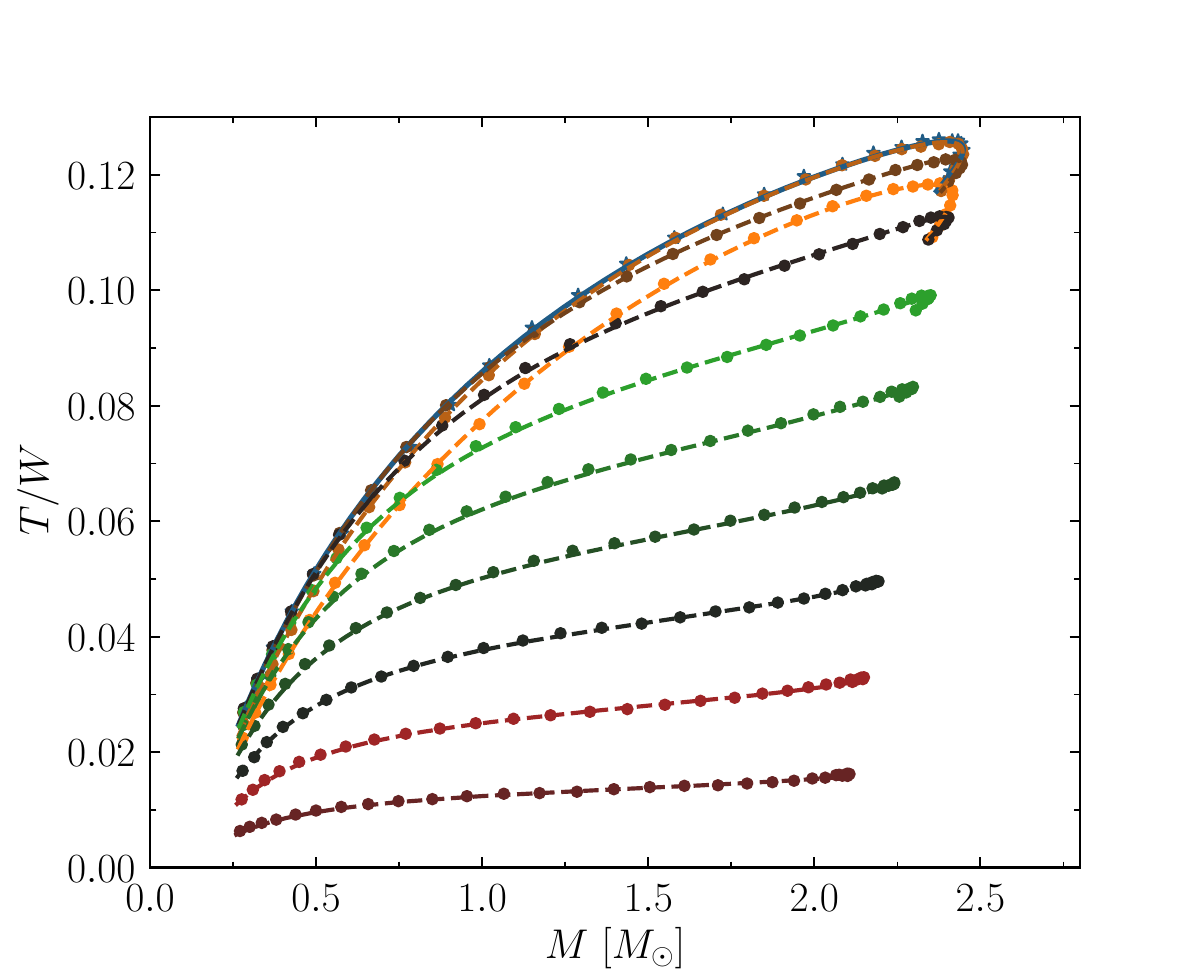}
\includegraphics[width=0.333\textwidth]{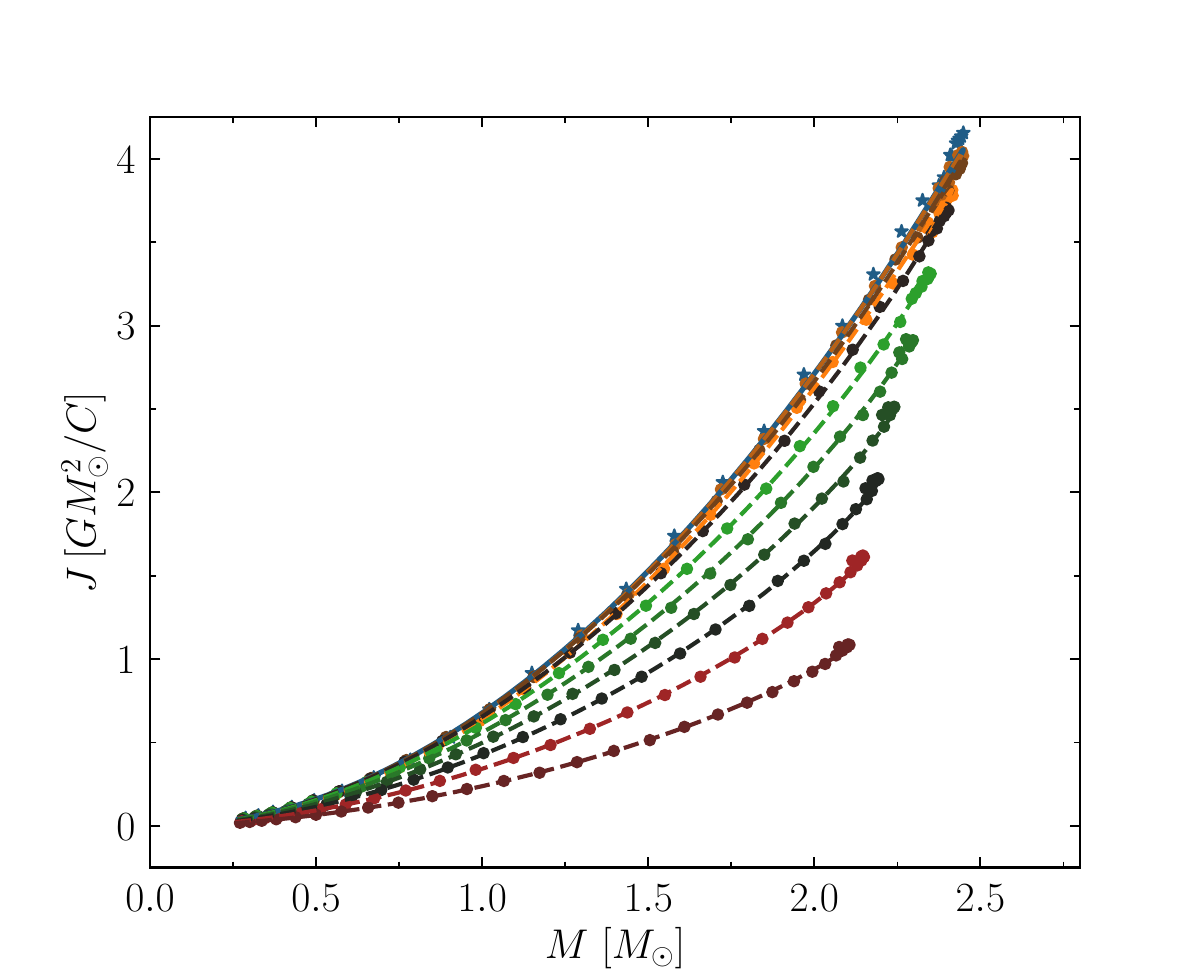}
\includegraphics[width=0.333\textwidth]{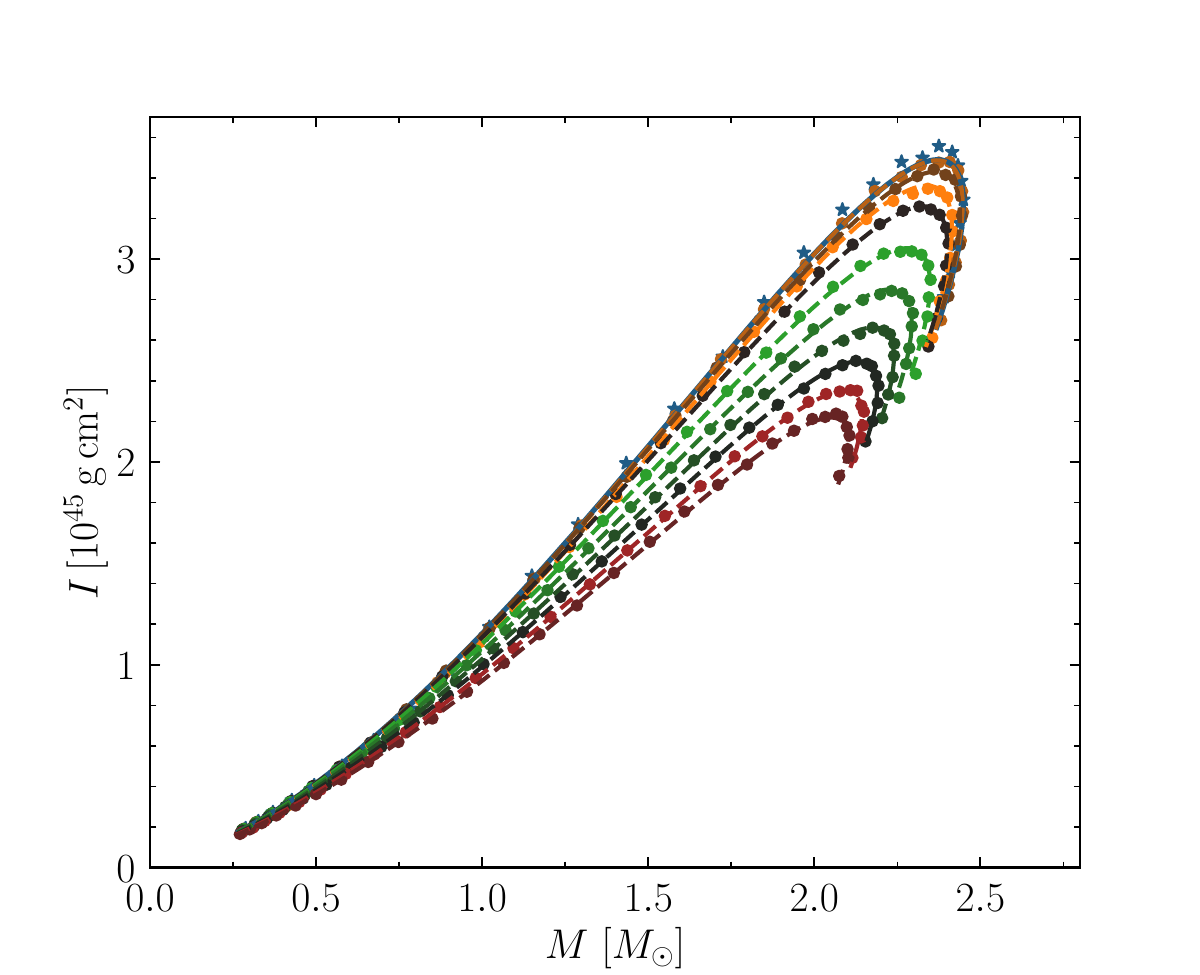}
\includegraphics[width=0.333\textwidth]{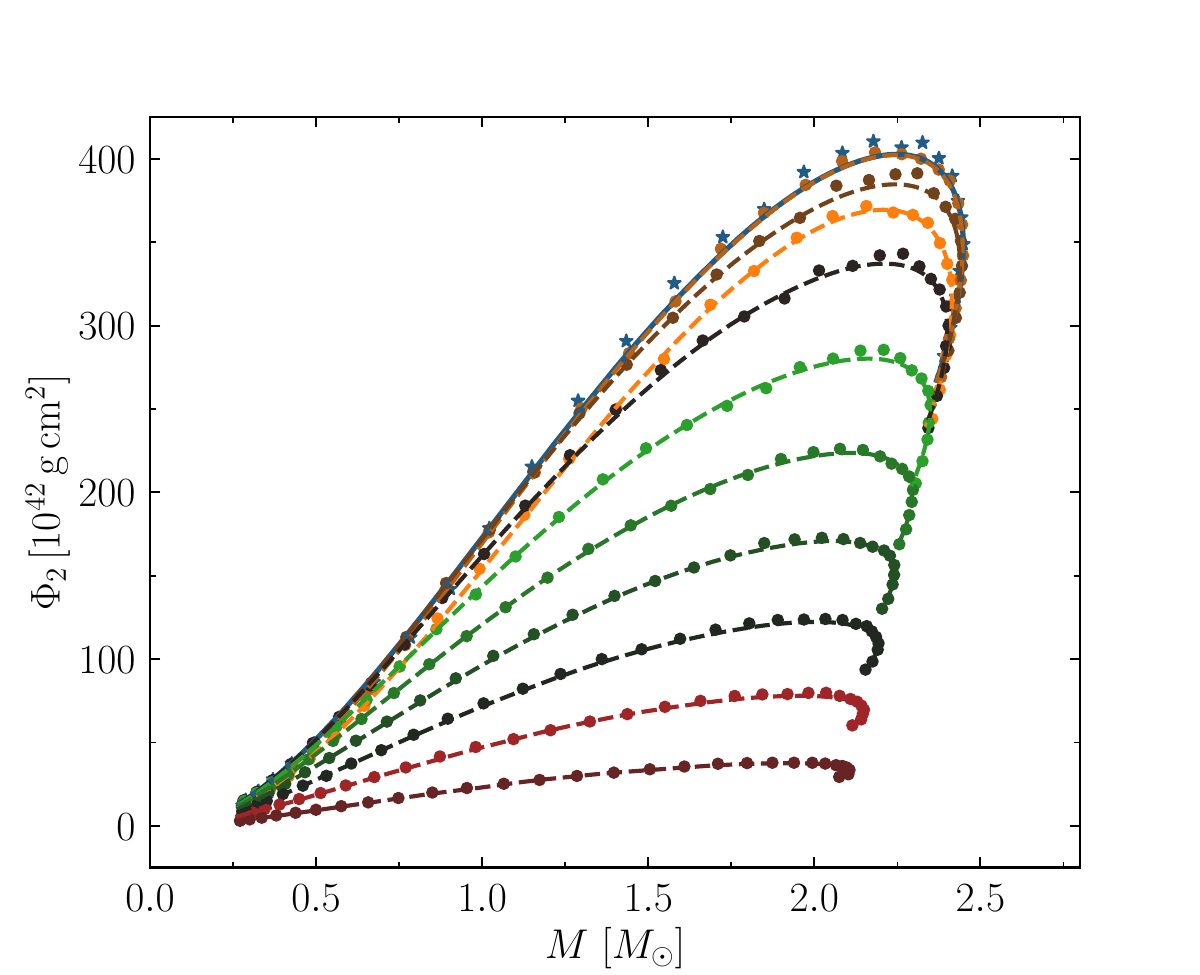}
\includegraphics[width=0.333\textwidth]{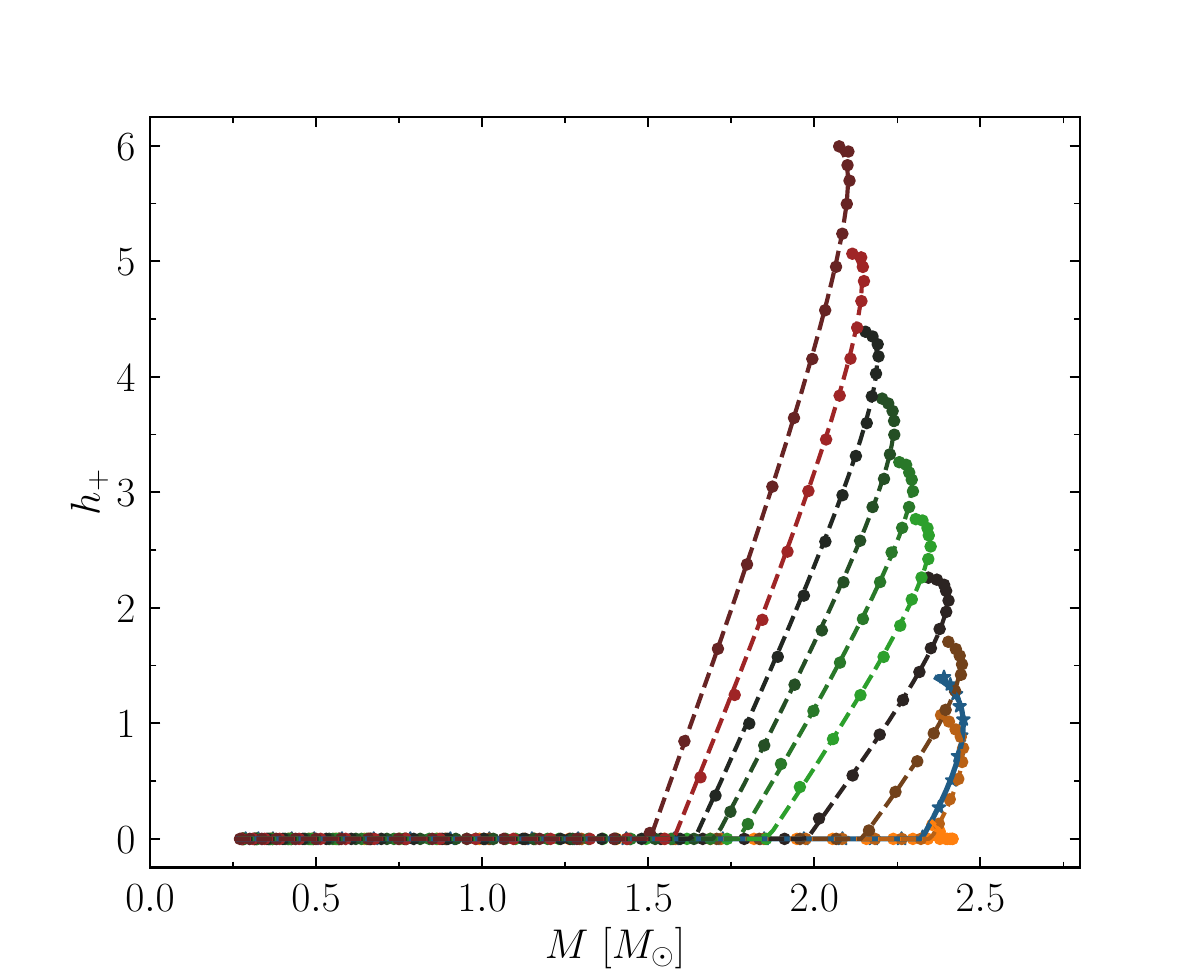}
\includegraphics[width=0.247\textwidth]{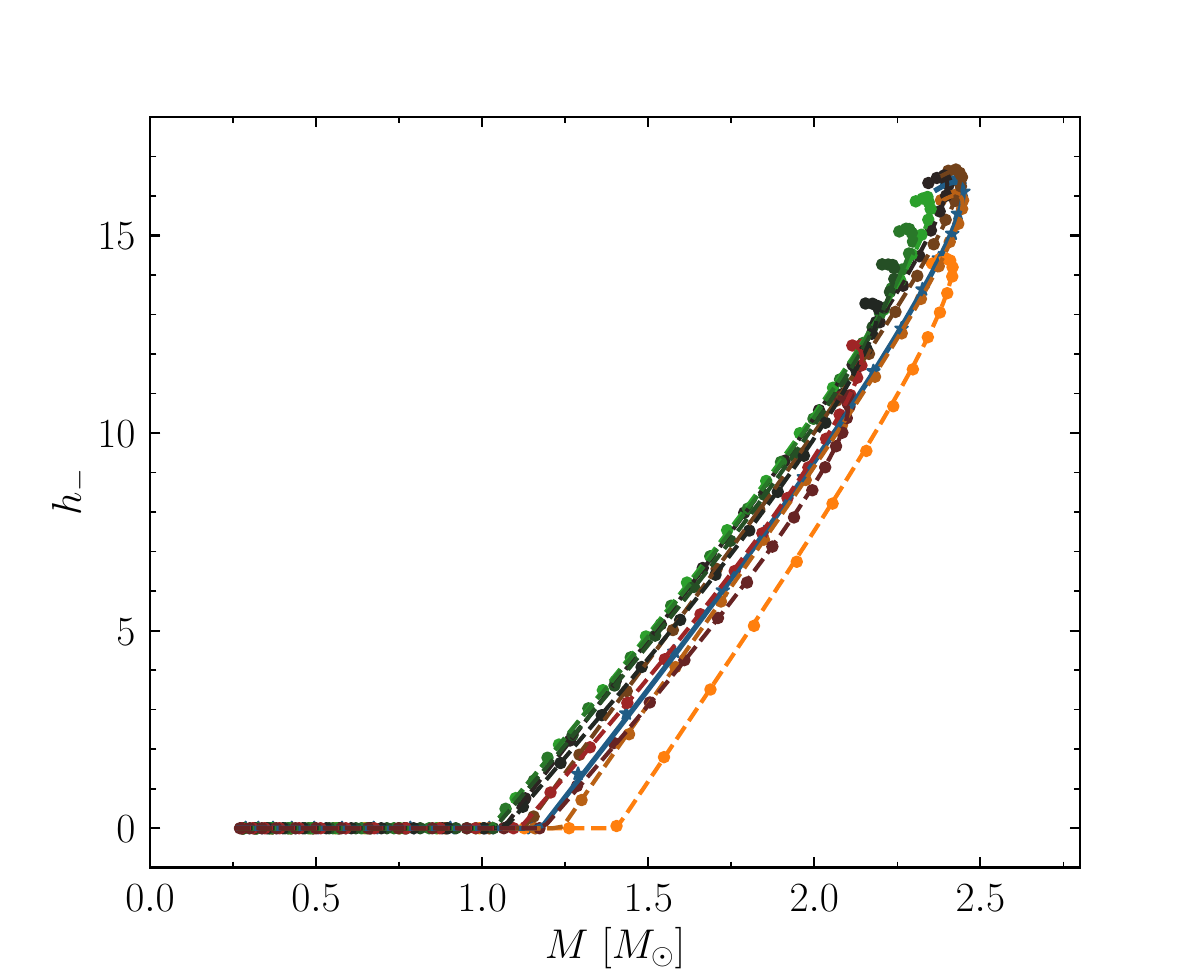}
\includegraphics[width=0.247\textwidth]{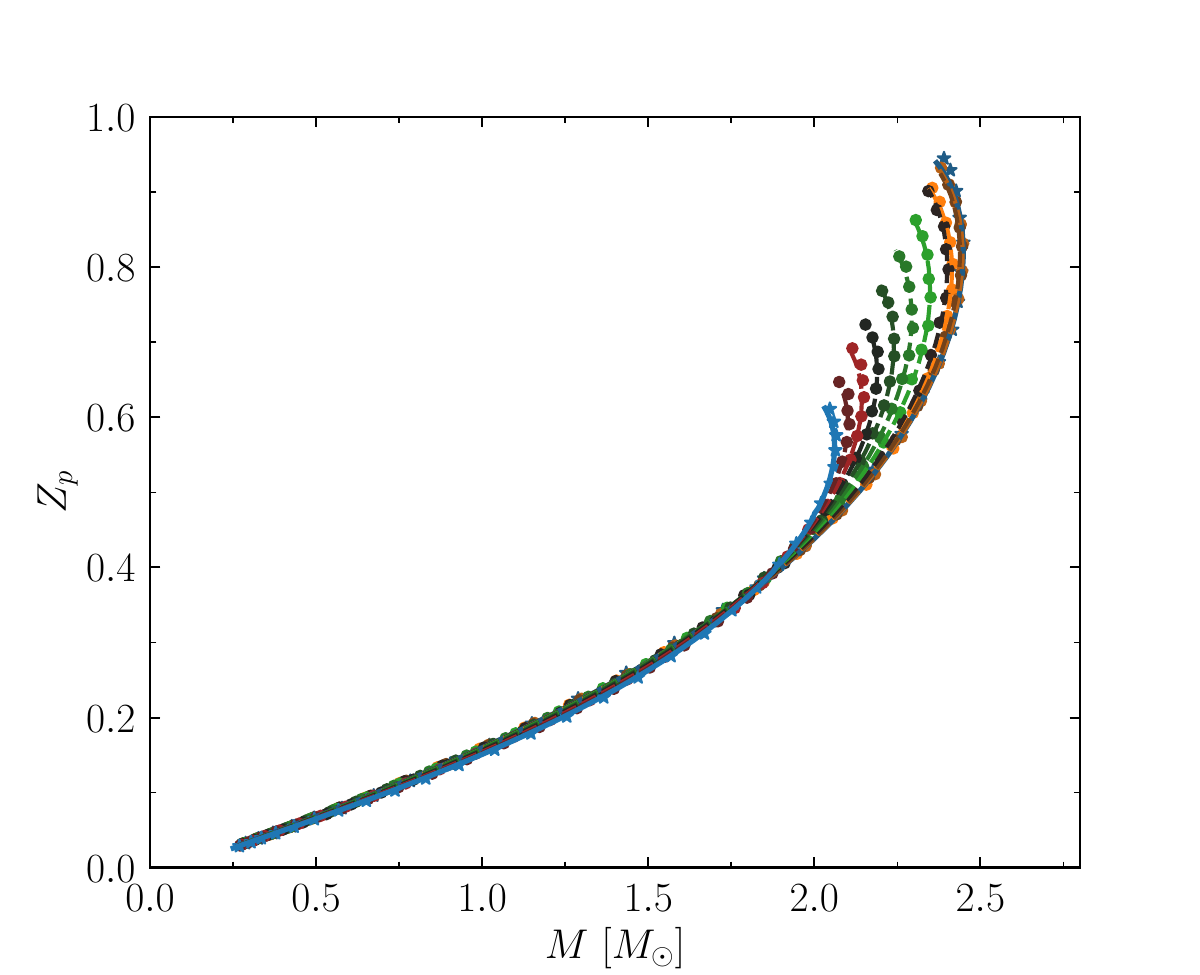}
\includegraphics[width=0.247\textwidth]{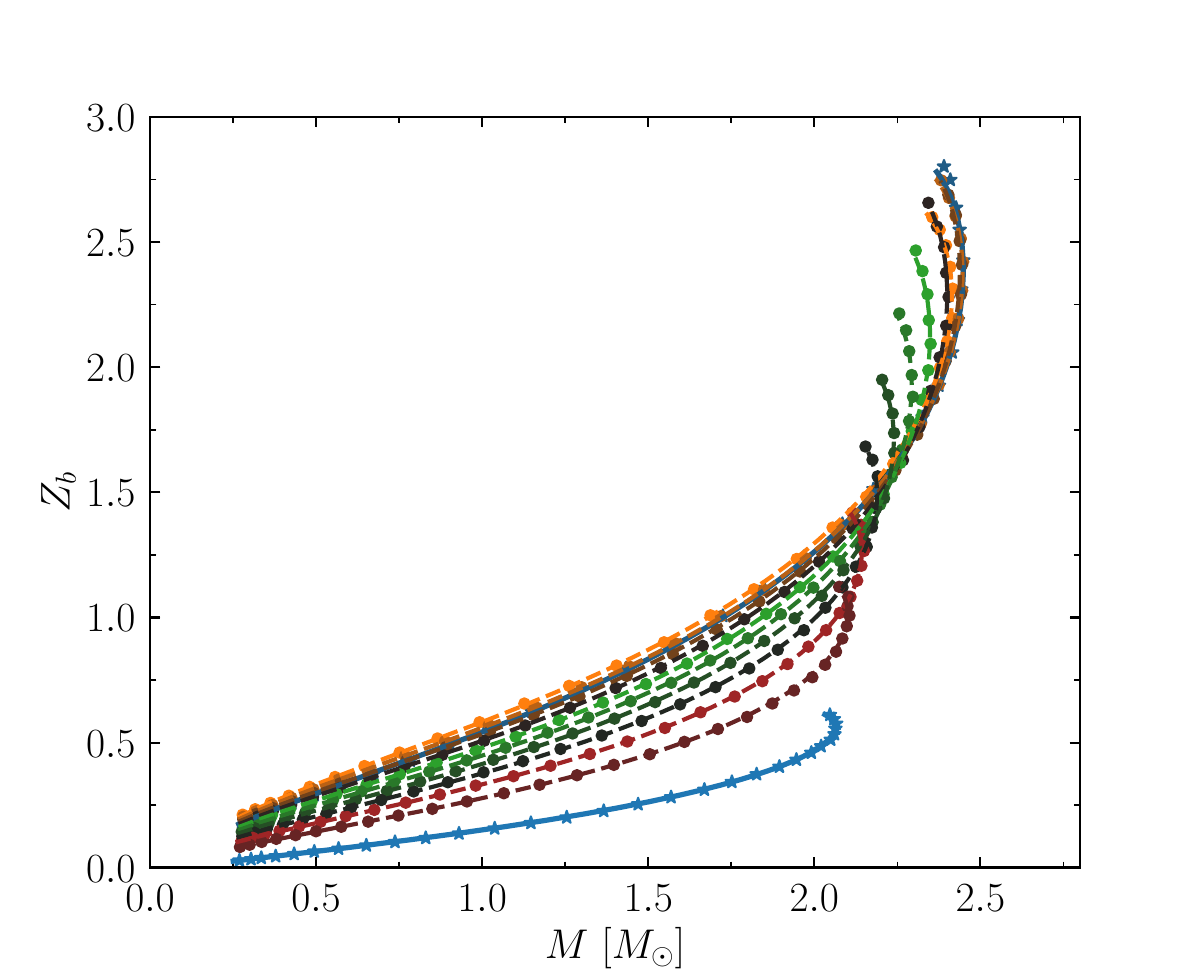}
\includegraphics[width=0.247\textwidth]{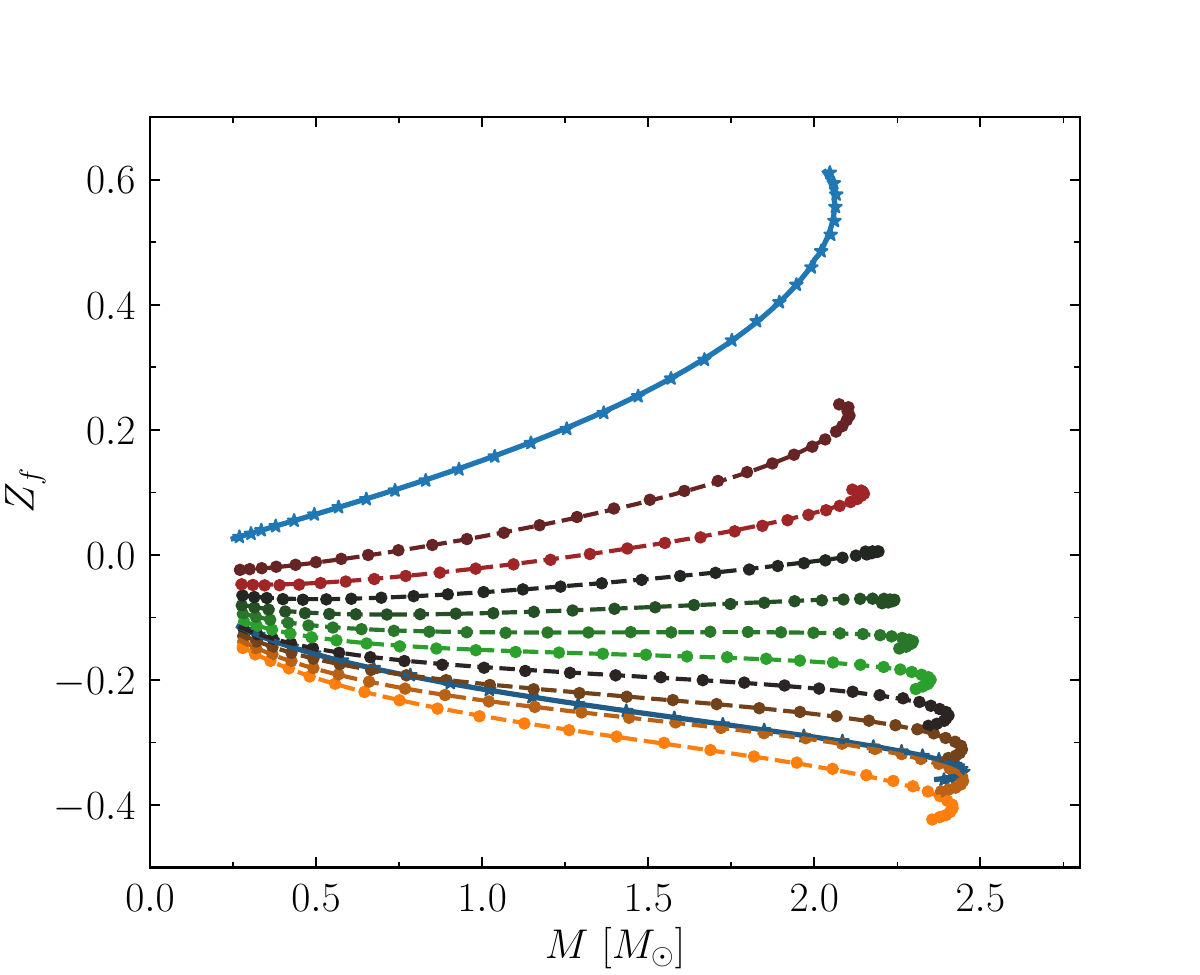}}
\caption{Other observables as a function of gravitational mass $M$ or radius $R$ for SFHo EoS. The labels are the same as in Fig.~\ref{fig:r_ratio_mr}}\label{fig:obs}
\vspace{-0.1cm}
\end{figure}

We show the comparison between the \texttt{RNS} ground truth and NN predictions for other observables in Fig.~\ref{fig:obs}. These observables include the baryonic mass $M_0$, rotational/gravitational energy $T/W$, angular momentum $J$, moment of inertia $I$, quadrupole moment $\Phi_2$, height of co-rotating innermost stable circular orbit (ISCO) $h_+$, height of counter-rotating ISCO $h_-$, polar redshift $Z_p$, backward equatorial redshift $Z_b$ and forward equatorial redshift $Z_f$. The predictions for all of these observables agree well with the ground truth.

\end{document}